\documentclass[10pt,a4paper]{article}

\usepackage[T1]{fontenc}
\usepackage[utf8]{inputenc}
\usepackage[a4paper,margin=1in]{geometry}
\usepackage{microtype}

\usepackage{mathtools}
\usepackage{amssymb}

\usepackage{amsthm}

\usepackage{graphicx}
\usepackage{booktabs}
\usepackage{multirow}
\usepackage{subcaption}

\usepackage{enumitem}
\usepackage[dvipsnames]{xcolor}

\usepackage[authoryear,round]{natbib}

\providecommand{\parencite}[1]{\citep{#1}}
\providecommand{\textcite}[1]{\citet{#1}}

\usepackage[
  colorlinks=true,
  linkcolor=blue,
  citecolor=blue,
  urlcolor=blue
]{hyperref}

\newtheorem{theorem}{Theorem}[section]

\newtheorem{proposition}[theorem]{Proposition}
\newtheorem{corollary}[theorem]{Corollary}

\theoremstyle{definition}
\newtheorem{definition}[theorem]{Definition}

\theoremstyle{remark}

\title{Exact and Approximate Range Queries in Ball Mapper}

\author{
Jay-Anne Bulauan\\
\small Independent Researcher\\
\small \href{mailto:bulauan_jayanne@yahoo.com}{bulauan\_jayanne@yahoo.com}
\and
John Rick Manzanares\\
\small Dioscuri Centre in Topological Data Analysis, Institute of Mathematics of the Polish Academy of Sciences\\
\small International Environmental Doctoral School of the University of Silesia in Katowice\\
\small \href{mailto:jdolormanzanares@impan.pl}{jdolormanzanares@impan.pl}
}

\date{\today}

\begin{document}
\maketitle

\begin{abstract}
Ball Mapper summarizes a finite metric dataset by covering the sample with closed balls centered at selected landmarks and connecting landmarks whose balls share observations. Its construction therefore depends critically on repeated fixed radius range queries, yet the effect of replacing exact queries by approximate search has not been systematically characterized. We formulate Ball Mapper through an abstract range query procedure that separates the mathematical construction from the search backend used to realize it. Under fixed ordering, exact procedures preserve the landmark sequence, cover, graph, and membership-based colorings. For approximate procedures, we derive deterministic bounds on covering radius and landmark separation under additive and multiplicative query errors, prove inclusions for the induced nerve, characterize edge survival through witness redundancy for conservative approximations, and bound perturbations of mean vertex colorings. The accompanying implementation provides independent exact reference backends together with exhaustive and approximate search methods under a common closed ball convention. Experiments on Gaussian, mixture, and noisy curve data across three seeds show that approximation fidelity depends strongly on geometry and that edges supported by multiple witnesses are substantially more robust to missed memberships. At 20,000 observations, the approximate indexes did not outperform exhaustive FAISS Flat search. The results therefore establish a framework for controlled approximation rather than a universal speed advantage, and identify the geometric and combinatorial quantities that govern when approximate range search preserves the Ball Mapper summary.
\end{abstract}

\begin{center}
\end{center}

\section{Introduction}

Ball Mapper is a method for constructing an interpretable graph summary of a finite metric dataset \parencite{dlotko2019ballmapper}. Given a scale parameter $\varepsilon>0$, the method selects representative data points, called landmarks, whose $\varepsilon$-balls cover the dataset. Each landmark becomes a vertex, and two vertices are connected when their corresponding balls contain a common data point. The resulting graph gives a coarse description of the organization of the data and can reveal clusters, branches, transitions, and regions with distinct values of an auxiliary variable.

The appeal of Ball Mapper lies partly in its metric formulation. It requires only a dataset, a distance function, and a scale parameter, and it does not assume a particular coordinate representation or probability model. This makes it suitable for exploratory analysis of structured and high-dimensional data. However, the same formulation also creates a computational bottleneck. During the greedy construction of the landmark set, each newly selected landmark requires a range query at a fixed radius to identify all data points lying within distance $\varepsilon$. A direct implementation performs this query by comparing the landmark with every point in the dataset.

The repeated range query computation can dominate the cost of Ball Mapper. The difficulty increases with the number of observations, because more distances must be evaluated, and with the ambient dimension, because each distance computation becomes more expensive. Moreover, geometric indexing methods can lose pruning efficiency in high dimensions, reflecting the broader curse of dimensionality \parencite{Bellman1957}. Thus, an efficient implementation of Ball Mapper should account for both the metric structure of the data and the computational cost of evaluating many distances.

The computational methods considered here fall into three broad classes: exact spatial indexes, exhaustive vector search, and approximate vector indexes. For exact spatial search, we consider ball trees and $k$-dimensional trees alongside a direct brute-force reference. A ball tree recursively partitions the dataset into bounded subsets and uses metric bounds to exclude subsets that cannot intersect a query ball \parencite{Omohundro1989}. A $k$-dimensional tree instead partitions the ambient coordinate space by recursive axis-aligned splits and prunes regions whose bounding boxes cannot intersect the query ball \parencite{Bentley1975}. These structures exploit different aspects of the data geometry, and their effectiveness depends on the extent to which the resulting partitions permit queries to discard observations without individual distance evaluations.

For dense Euclidean data, we also consider Facebook AI Similarity Search (FAISS), a library for efficient vector search \parencite{douze2025faisslibrary}. A Flat index performs exhaustive search: every database vector remains a candidate, but distance computations are implemented using optimized vectorized kernels, multithreading, and cache-conscious memory access. It therefore provides a useful exhaustive backend against which indexed methods can be compared. Because the FAISS computations use finite-precision vector representations, however, points lying very close to the prescribed radius require particular care when exact agreement with the closed-ball definition is required.

We further consider non-exhaustive FAISS indexes based on inverted-file and hierarchical navigable small-world search. An inverted-file (IVF) index partitions the dataset into coarse cells and searches only selected cells, whereas a hierarchical navigable small-world (HNSW) index explores a multilevel proximity graph to locate promising regions of the dataset. Both methods can reduce the number of observations examined by a query, but they may also omit points that belong to the exact query ball. From the perspective of Ball Mapper, such omissions are consequential: they can change ball memberships, alter the greedy landmark sequence, remove witnesses of intersections, and thereby modify the resulting graph and its membership-based colorings.

The effect of approximate search on Ball Mapper requires a range query analysis rather than a conventional nearest neighbor analysis. Ball Mapper requires the set of all observations lying within a prescribed radius, not only a sufficiently close observation or a fixed number of nearest neighbors. Approximation is therefore described through inclusions between exact and approximate range sets. Under additive and multiplicative range errors, we obtain deterministic bounds on the covering radius and minimum separation of the resulting landmark set. For a fixed landmark set, we further derive inclusions between approximate and exact nerves at nearby scales, characterize the survival of graph edges through witness redundancy, and bound perturbations of membership-based vertex colorings.

The empirical part of the paper has three purposes. First, we compare ball tree and exhaustive FAISS Flat implementations of the range query step against the existing \emph{pyBallMapper} implementation \parencite{pyballmapper}. Second, we validate the closed ball convention and the resulting Ball Mapper construction against independent double precision floating point reference and exact spatial backends, including brute force, ball tree, and $k$-dimensional tree. Third, we study the structural effects of approximate range search using IVF and HNSW. The approximation experiments focus not primarily on speed, but on fidelity of ball memberships, landmark selection, witness counts, graph edges and connectivity, and membership based vertex colorings.

We use the standard mathematical and computer science term \emph{oracle} for an idealized procedure that returns the answer to a specified query. In this article, a \emph{range query oracle} returns all observations lying in a requested metric ball. A search backend is a concrete computational realization of such an oracle. This terminology separates the mathematical range set from the data structure, numerical representation, or software library used to compute it.

The main contributions of this work are as follows:
\begin{enumerate}
\item We formulate Ball Mapper construction through an explicit range query oracle and prove that, under fixed observation ordering and tie breaking, any exact backend produces the same landmark sequence, cover, graph, and membership based colorings.

\item We analyze the computational roles of exhaustive and tree based range search
and benchmark BallTree and exhaustive FAISS Flat implementations against \emph{pyBallMapper}.

\item For additive and multiplicative range errors, we derive deterministic bounds on the covering radius and minimum landmark separation. For a fixed landmark set, we further establish inclusions between approximate and exact nerves at nearby scales, with corresponding consequences for the Ball Mapper graph.

\item For conservative approximate search, we characterize edge survival in terms of witness redundancy and derive a perturbation bound for mean vertex colorings.

\item We show empirically that sensitivity to approximate range search depends strongly on data geometry and witness redundancy. At the tested scale of 20,000 observations, the approximate indexes do not provide a runtime advantage over exhaustive FAISS Flat search.
\end{enumerate}

The objective is not to alter the definition of Ball Mapper, but to separate the mathematical construction from the computational mechanism used to realize its metric balls and to make the consequences of approximate range search mathematically explicit.

\section{Ball Mapper}

This section introduces the Ball Mapper. We first present the theoretical background underpinning the method, then describe the construction of the cover and associated graph. Finally, we discuss the coloring schemes used to encode and analyze auxiliary data.

\subsection{Cover Construction}

To describe the Ball Mapper algorithm rigorously, we first outline the mathematical setting in which it operates. First, we model the dataset as a metric space. A \emph{metric space} $(X,d)$ is a set $X$ together with a \emph{distance function} $d:X \times X \rightarrow \mathbb{R}$ satisfying, for all $x,y,z \in X$,

\begin{enumerate}
\item
  $d(x,y) \geq 0$,
\item
  $d(x,y) = 0$ if and only if $x = y$,
\item
  $d(x,y) = d(y,x)$, and
\item
  $d(x,z) \leq d(x,y) + d(y,z)$.
\end{enumerate}

Given a metric structure on the dataset, the next step is to select a finite set of representative points that captures the geometry of the space at a prescribed scale. This is achieved through the notion of an $\varepsilon$-net.

\begin{definition}
Let $(X,d)$ be a metric space. In this article, for a real number $\varepsilon > 0$, an \emph{$\varepsilon$-net} of $X$ means a subset $N \subset X$ satisfying the closed ball convention used throughout this work:
\begin{enumerate}
\item
  For every point $x \in X$, there exists $n \in N$ with
  $d(x,n) \leq \varepsilon$.
\item
  For any pair of distinct points $n_{1},n_{2} \in N$, we have
  $d\left( n_{1},n_{2} \right) > \varepsilon$.
\end{enumerate}

Elements of $N$ are called \emph{landmarks}.
\end{definition}

An $\varepsilon$-net naturally induces a collection of neighborhoods by associating to each landmark a ball of radius $\varepsilon$. Specifically, an $\varepsilon$-net $\left\{ c_{1},c_{2},\ldots,c_{k} \right\}$ yields a collection of \emph{closed balls} of radius $\varepsilon$
\[
B\left( c_{i},\varepsilon \right) = \left\{ y \in X~|~d(c_i,y) \leq \varepsilon \right\}.
\]
centered at the landmarks $c_{i}$. The union of these balls is said to \emph{cover} the metric space $X$.

The parameter $\varepsilon$ controls the scale at which the metric structure of the dataset is examined. Smaller values of $\varepsilon$ produce covers with more balls of smaller radius, yielding a finer representation of local geometric features. Larger values result in coarser covers in which local variations are aggregated. In practice, $\varepsilon$ is chosen to balance resolution against robustness to sampling noise.

Given a cover of the dataset, a standard way to encode the pattern of overlaps among the covering sets is through an abstract simplicial complex. This construction records not only pairwise intersections but also higher-order overlaps in a combinatorial form.

An \emph{(abstract) simplicial complex} is a pair $(V,S)$ where $V$ is a set of \emph{vertices}, and $S$ is a collection of nonempty finite subsets of $V$ satisfying the following:
\begin{enumerate}
\item
  For every $v \in V$, $\left\{ v \right\} \in S$.
\item
  If $\sigma \in S$ and $\tau \subset \sigma$, then $\tau \in S$.
\end{enumerate}
Elements $\sigma$ of $S$ are called \emph{$k$-simplices} where the \emph{dimension} $k = |\sigma| - 1$.

\begin{definition}
The \emph{nerve} of the cover induced by an $\varepsilon$-net is the abstract simplicial complex defined as follows. The vertex set consists of the landmarks $\left\{ c_{1},\ldots,c_{k} \right\}$. A finite subset $\{ c_{i_{0}},\ldots,c_{i_{m}}\}$ of landmarks spans an $m$-simplex if the corresponding closed balls of radius $\varepsilon$ have a nonempty common intersection, that is,
\[
\bigcap_{j = 0}^{m}B\left( c_{i_{j}},\varepsilon \right) \cap X \neq \varnothing.
\]
\end{definition}

A \emph{$k$-skeleton} of a simplicial complex $S$ is a simplicial complex containing all simplices of $S$ whose dimension is at most $k$. The \emph{Ball Mapper graph} is the $1$-skeleton of the nerve of an $\varepsilon$-net. The $0$-simplices (or \emph{vertices}) are precisely the landmarks, while the $1$-simplices (or \emph{edges}) between two vertices are formed when the associated landmarks have a nonempty intersection.

While higher-order simplices encode multiway intersections among balls, restricting attention to the $1$-skeleton yields a graph representation that is computationally efficient and easy to interpret. This representation preserves the dominant pairwise connectivity structure and is well-suited to exploratory analysis and visualization.

The classical Nerve Lemma \parencite{Borsuk1948} identifies conditions under which the nerve of a cover has the homotopy type of the covered space. Ball Mapper instead forms a cover of the observed finite dataset and retains only the $1$-skeleton. We therefore do not invoke the Nerve Lemma as a homotopy-recovery guarantee here; the Ball Mapper graph is treated as a multiscale descriptive summary of observed overlap structure.

\subsection{Practical Implementation}

Ball Mapper constructs a cover of a finite dataset $X$ by computing an $\varepsilon$-net using a greedy algorithm. The procedure is as follows:

\begin{enumerate}
\item
  Initialize an empty set $N$ to hold the landmarks.
\item
  While there exist uncovered points in the dataset:
  \begin{itemize}
  \item
    Select an uncovered point $x$ from the dataset.
  \item
    Add $x$ to $N$.
  \item
    Mark all points within distance $\varepsilon$ of $x$ as covered.
  \end{itemize}
\end{enumerate}

A point is said to be \emph{covered} if it lies within at least one ball centered at a landmark in $N$; otherwise, it is \emph{uncovered}.

\begin{proposition}\label{prop:greedy-maximal}
Assume that the greedy procedure marks precisely the points in the closed range set $R_\varepsilon(l)=\{x\in X:d(x,l)\leq\varepsilon\}$ whenever a landmark $l$ is selected. Then the final landmark set $L$ is $\varepsilon$-separated under the strict convention $d(l_i,l_j)>\varepsilon$, and is maximal with respect to this property. Consequently, $L$ is an $\varepsilon$-cover of $X$.
\end{proposition}

\begin{proof}
If $l_i$ is selected before $l_j$, then $l_j$ is still uncovered when it is selected. Hence, $d(l_i,l_j)>\varepsilon$, which proves separation. At termination, every point not in $L$ has been covered by a selected landmark. Thus, for every $x\in X\setminus L$, there exists $l\in L$ such that $d(x,l)\leq\varepsilon$. Therefore, no further point can be added to $L$ while preserving $\varepsilon$-separation, and the same statement gives the $\varepsilon$-cover property.
\end{proof}

This proposition makes explicit that the greedy Ball Mapper landmarks form an inclusion-maximal $\varepsilon$-separated set and simultaneously an $\varepsilon$-cover. The particular maximal set depends on the processing order, but the separation and covering conclusions do not.

This greedy construction depends on the order in which points are processed, and consequently the resulting set of landmarks is not unique. 
An alternative approach is \emph{farthest point sampling} \parencite{GONZALEZ1985293}. It can be made deterministic after fixing the initial landmark and a deterministic rule for resolving distance ties; without these choices, the resulting landmark sequence need not be unique. The procedure is as follows:

\begin{enumerate}
\item
  Select an initial landmark $x_{0} \in X$.
\item

  Initialize the current landmark set as $L=\{x_0\}$.

\item
  Find a point
  \[x^{\ast} = \arg\max\limits_{x \in X}\min\limits_{l \in L}d(x,l)\]
  farthest from the current landmark set.
\item

  Add $x^{\ast}$ to $L$ if
  $\min\limits_{l \in L}d\left( x^{\ast},l \right) > \varepsilon$.

\item
  Repeat until no such point exists.
\end{enumerate}

For a fixed initial landmark and deterministic tie-breaking rule, this procedure yields a reproducible landmark sequence. Different initial landmarks or tie resolutions can nevertheless produce different $\varepsilon$-nets. The strict condition in Step 4 is consistent with the closed ball convention used above: a point at distance exactly $\varepsilon$ is already covered and is therefore not eligible to become a new landmark. The procedure is typically more computationally expensive than the one-pass greedy construction and is therefore less commonly used in large-scale applications.

\subsection{Coloring}

In applications, the Ball Mapper is typically used as a scaffold for visualizing auxiliary information associated with the data, such as response variables, labels, or summary statistics. By aggregating such quantities over balls, one can examine how they vary across regions of the data space while respecting its metric structure.

Let $M$ be a Ball Mapper graph with landmark set $L$. A \emph{coloring} of $M$ is a function $c:L \rightarrow \mathbb{R}$ where $c(l)$ encodes aggregated information associated with the closed ball centered at $l \in L$.

Let $f:X \rightarrow \mathbb{R}$ be a real-valued function defined on the dataset $X$. For example, $f$ can be some response variable or feature. Given a landmark $l_{i}$ of $M$, let $X_{i} = X \cap B\left( l_{i},\varepsilon \right)$. A common choice of coloring is given by averaging:
\[
c\left( l_{i} \right) = \frac{1}{|X_{i}|}\sum_{x \in X_{i}}f(x).
\] 
This assigns to each vertex a summary statistic of the data points contained in its corresponding ball.

To understand how coloring behaves across the graph, we impose a mild regularity assumption on $f$. Note that $\mathbb{R}$ can be viewed as a metric space with the Euclidean distance function $| \cdot |$.

\begin{definition}
The function $f$ is said to be \emph{$k$-Lipschitz continuous} if, for some constant $k$, 
\[
|f(x) - f(y)| \leq kd(x,y)
\]
for all $x,y \in X$.
\end{definition}

For a Lipschitz continuous function, the difference in the images of any two points is at most proportional to the distance between the points.

The preceding Lipschitz assumption permits a precise analysis of colorings aggregated over balls, but the bound depends on how the color is formed. To include the mean, weighted mean, median, trimmed mean, minimum, and maximum in one framework, it is convenient to use convex aggregation.

\begin{definition}
For a nonempty ball membership set $X_i$, a \emph{convex ball coloring} is any quantity of the form
\[
c(l_i)=\sum_{x\in X_i}\alpha_{i,x}f(x).
\]
The weights satisfy $\alpha_{i,x}\geq0$ for every $x\in X_i$, and $\sum_{x\in X_i}\alpha_{i,x}=1$. The ordinary mean is obtained with $\alpha_{i,x}=|X_i|^{-1}$. Because $X_i$ is finite, the minimum and maximum correspond to placing unit weight on a minimizing or maximizing point, and a median or trimmed mean can likewise be represented by suitable nonnegative weights.
\end{definition}

\begin{proposition}\label{prop:centerColor}
Let $f:X\to\mathbb R$ be $k$-Lipschitz and let $c$ be a convex ball coloring. Then
\[
|c(l_i)-f(l_i)|\leq k\varepsilon.
\]
\end{proposition}

\begin{proof}
Since $l_i\in X_i$ and $d(x,l_i)\leq\varepsilon$ for every $x\in X_i$, Lipschitz continuity gives $|f(x)-f(l_i)|\leq k\varepsilon$. Therefore,
\[
|c(l_i)-f(l_i)|
=\left|\sum_{x\in X_i}\alpha_{i,x}(f(x)-f(l_i))\right|
\leq\sum_{x\in X_i}\alpha_{i,x}|f(x)-f(l_i)|
\leq k\varepsilon.
\]
\end{proof}

\begin{corollary}\label{cor:localVariation}
Under the assumptions of Proposition~\ref{prop:centerColor}, for every $x\in X_i$,
\[
|f(x)-c(l_i)|\leq2k\varepsilon.
\]
\end{corollary}

\begin{proof}
By the triangle inequality,
\[
|f(x)-c(l_i)|\leq |f(x)-f(l_i)|+|f(l_i)-c(l_i)|\leq2k\varepsilon.
\]
\end{proof}

The factor $2$ is necessary for a general aggregated coloring. Although every point in $X_i$ lies within $\varepsilon$ of the landmark, two points in the same ball can be almost $2\varepsilon$ apart. Thus, a bound of $k\varepsilon$ for $|f(x)-c(l_i)|$ is not valid without an additional restriction such as choosing $c(l_i)=f(l_i)$.

\begin{proposition}\label{prop:adjacentColor}
Let $f$ be $k$-Lipschitz and let $c$ be a convex ball coloring. If $(l_i,l_j)$ is an edge of the Ball Mapper graph, then
\[
|c(l_i)-c(l_j)|\leq4k\varepsilon.
\]
If the landmark value coloring $c(l)=f(l)$ is used instead, the sharper bound
\[
|c(l_i)-c(l_j)|\leq2k\varepsilon
\]
holds.
\end{proposition}

\begin{proof}
Let $x\in X_i\cap X_j$ be a witness for the edge. Then
\[
d(l_i,l_j)\leq d(l_i,x)+d(x,l_j)\leq2\varepsilon.
\]
For a convex ball coloring, Proposition~\ref{prop:centerColor} and Lipschitz continuity give
\begin{align*}
|c(l_i)-c(l_j)|
&\leq |c(l_i)-f(l_i)|+|f(l_i)-f(l_j)|+|f(l_j)-c(l_j)|\\
&\leq k\varepsilon+2k\varepsilon+k\varepsilon
=4k\varepsilon.
\end{align*}
If $c(l)=f(l)$, only the middle term remains.
\end{proof}

\begin{corollary}\label{cor:pathColor}
If the graph distance between $l_i$ and $l_j$ is $r<\infty$, then every convex ball coloring generated by a $k$-Lipschitz function satisfies
\[
|c(l_i)-c(l_j)|\leq2k\varepsilon(r+1).
\]
For landmark value coloring, the bound improves to $2k\varepsilon r$.
\end{corollary}

\begin{proof}
A path of $r$ edges gives $d(l_i,l_j)\leq2r\varepsilon$ by repeated use of the triangle inequality through edge witnesses. Applying Proposition~\ref{prop:centerColor} at the two endpoints gives
\[
|c(l_i)-c(l_j)|\leq k\varepsilon+2rk\varepsilon+k\varepsilon
=2k\varepsilon(r+1).
\]
For landmark value coloring, Lipschitz continuity gives the sharper bound directly.
\end{proof}

These results formalize two distinct approximation statements. The vertex color approximates the value of $f$ at its landmark to order $\varepsilon$, while the graph controls how rapidly colors can change along edges and paths. Statistics such as variance and range are not convex aggregations on the original scale of $f$ and therefore require separate bounds.

When $f$ is not Lipschitz, the preceding deterministic variation bounds no longer apply. Sharp color transitions may then occur, and the Ball Mapper graph can be used heuristically to identify regions of qualitatively different behavior. This may happen, for example, when $f$ represents class labels in classification problems. In such cases, robust or categorical summaries such as the mode or majority vote are often more appropriate than averaging-based colorings.

In both perspectives, the choice of coloring determines the role played by the Ball Mapper. Smooth colorings emphasize geometric continuity, while nonsmooth colorings expose boundaries and regimes. Neither viewpoint is preferable a priori; each answers a different question about the data.

Lastly, the coloring depends on the choice of landmarks. The greedy $\varepsilon$-net construction is order-dependent and may yield non-unique landmark sets, resulting in non-consistent colorings. Farthest point sampling reduces this source of variability only after its initial landmark and tie-breaking rule are fixed; under those additional choices it is reproducible, but it is not uniquely determined by the data and $\varepsilon$ alone.

\section{Range Query Backends}\label{sectionQuery}

Constructing an $\varepsilon$-net requires repeatedly identifying all points within distance $\varepsilon$ of a selected landmark. This operation is a \emph{range query}. It is useful to regard the Ball Mapper algorithm as using a range query oracle independently of the underlying search data structure. The mathematical algorithm asks only for the set $R_\varepsilon(q)$; brute force, metric trees, vector search libraries, or approximate candidate generators are computational realizations of that oracle. This separation is central to the analysis below: exact backends may change runtime but not the Ball Mapper object, whereas approximate backends change the oracle itself and therefore require stability guarantees.

In the following theorem, we fix the data order, the scale $\varepsilon$, and all tie-breaking rules used by the Ball Mapper implementation.

\begin{theorem}\label{thm:exact-oracle-invariance}
Suppose two implementations return exactly the same range set $R_\varepsilon(q)$ for every query point $q$ encountered during construction. Then they select the same landmarks in the same order, produce identical ball memberships, and therefore produce the same nerve, Ball Mapper graph, and any coloring computed solely from those memberships.
\end{theorem}

\begin{proof}
Let $U_k^{A}$ and $U_k^{B}$ denote the sets of uncovered observations after $k$ landmark selections for implementations $A$ and $B$, respectively. We prove by induction on $k$ that the two implementations have selected the same first $k$ landmarks and that
\[
U_k^{A}=U_k^{B}.
\]

For $k=0$, no landmarks have been selected and no observations have been covered, so
\[
U_0^{A}=U_0^{B}=X.
\]

Suppose the claim holds after $k$ selections. Since $U_k^{A}=U_k^{B}$, the common deterministic selection rule chooses the same next landmark $q$ in both implementations. By hypothesis, both implementations return exactly the same range set $R_\varepsilon(q)$. Consequently,
\[
U_{k+1}^{A} = U_k^{A}\setminus R_\varepsilon(q) = U_k^{B}\setminus R_\varepsilon(q) = U_{k+1}^{B}.
\]
Thus the $(k+1)$-st landmark is the same and the uncovered sets remain equal. The induction therefore gives the same landmark sequence for the entire construction.

For each selected landmark, the two implementations also return the same range set, so their ball memberships are identical. The nerve is determined entirely by intersections among these membership sets, and the Ball Mapper graph is its $1$-skeleton. Hence, both are identical. Any coloring determined solely by the membership sets and fixed observation data is likewise identical.
\end{proof}

Thus, an exact acceleration changes only the computational route, not the mathematical Ball Mapper object. In floating-point implementations, this statement should be interpreted relative to a fixed numerical comparison rule at the radius boundary.

The associated implementation in this study represents the mathematical closed ball predicate $d(x,l)\leq\varepsilon$ using an explicit floating-point boundary convention. Let $\mathbb{F}$ denote the set of finite numbers representable in the floating-point format used by a given comparison, and let $\varepsilon\in\mathbb{F}$. Define
\[
\varepsilon^+ = \min\{r\in\mathbb{F}:r>\varepsilon\},
\]
that is, the smallest representable floating-point number strictly larger than $\varepsilon$. In the implementation, this adjacent representable value is obtained by a standard \texttt{nextafter} operation directed toward $+\infty$. Since no representable floating-point number lies strictly between $\varepsilon$ and $\varepsilon^+$, a strict comparison of the form
\[
d(x,l)<\varepsilon^+
\]
is equivalent, at the level of the represented values, to the inclusive comparison $d(x,l)\leq\varepsilon$.

For ordinary radius comparisons, the implementation constructs $\varepsilon^+$ in double precision. FAISS Euclidean range search instead operates on single-precision squared Euclidean distances and uses a strict threshold. Accordingly, the implementation first represents $\varepsilon^2$ in single precision and then replaces that value by the next larger representable single-precision number before issuing the FAISS range query. Exact backends are subjected to an explicit closed ball filtering step so that the same mathematical membership criterion is enforced throughout. 

This outward rounding is a device for representing a non-strict boundary with an interface that uses a strict floating-point comparison. It does not change the intended mathematical radius. It also does not remove numerical error in the computed distances themselves, and therefore does not by itself guarantee agreement between computations performed in different floating-point precisions.

\subsection{Exact Backend Implementations}

The implementation uses complementary exact strategies. Scikit-learn BallTree accelerates metric range queries by pruning tree nodes whose metric bounds cannot intersect the query ball. FAISS Flat instead performs an exhaustive scan over dense vectors but reduces the constant cost of distance evaluation through optimized vector and linear algebra kernels. The double-precision floating-point brute force backend serves as the transparent reference oracle, while SciPy cKDTree provides an independent exact Euclidean tree implementation \parencite{Virtanen2020SciPy}. The Ball Mapper construction is independent of these choices; only the backend determines the available metric, numerical representation, and computational strategy. Additional implementation details are given in Appendix~\ref{app:backend-details}.

\subsection{Range Query Complexity}

Let $X=\{x_1,\ldots,x_n\}\subseteq\mathbb R^D$ be the dataset. Let $L\subseteq X$ denote the set of landmarks selected by the greedy Ball Mapper construction, and let $m=|L|$ be the number of selected landmarks. Since the landmarks are observations from $X$, the inequality $m\leq n$ always hold.

For a landmark $l\in L$, a radius query asks for all observations whose Euclidean distance from $l$ does not exceed the prescribed radius $\varepsilon$. The Euclidean distance between $x=(x_1,\ldots,x_D)$ and $l=(l_1,\ldots,l_D)$ is
\[
d(x,l) = \left(\sum_{j=1}^{D}(x_j-l_j)^2\right)^{1/2}.
\]
A \emph{point-to-point distance evaluation} means computing this distance, or an equivalent quantity such as its square, for one pair of observations. Computing a Euclidean distance requires examining all $D$ coordinates. Its arithmetic cost therefore grows proportionally to $D$.

We use standard asymptotic notation to describe how computational cost grows as the problem size increases. For two nonnegative quantities $f$ and $g$, the notation $f=O(g)$ means that, apart from a constant factor, $f$ is eventually no larger than $g$. Thus $O(g)$ gives an asymptotic upper bound. The notation $f=\Theta(g)$ means that $f$ is bounded both above and below by constant multiples of $g$ for sufficiently large problem sizes. Thus $\Theta(g)$ describes the order of growth rather than only an upper bound. The constants suppressed by this notation do not depend on the problem-size parameters under consideration.

Since one range query is issued for each selected landmark, the cost of the range query stage depends on both the number $n$ of observations and the number $m$ of landmarks.

\begin{proposition}\label{prop:flat-complexity}
An exhaustive implementation evaluates the distance from each of the $m$ landmarks to all $n$ observations. It therefore performs exactly $mn$ point-to-point distance evaluations, and hence $\Theta(mn)$ such evaluations. For dense Euclidean vectors in $\mathbb R^D$, each distance evaluation requires $\Theta(D)$ arithmetic work, so the total arithmetic cost is $\Theta(mnD)$. Since $m\leq n$, this cost is at most $O(n^2D)$.
\end{proposition}

\begin{proof}
Each of the $m$ selected landmarks generates one exhaustive range query. For a fixed landmark, an exhaustive query compares that landmark with all $n$ observations, giving $n$ distance evaluations. Repeating this for all $m$ landmarks gives $mn$ evaluations.

For Euclidean data in $\mathbb R^D$, evaluating the distance between two points requires processing their $D$ coordinate pairs. The amount of arithmetic per distance is therefore proportional to $D$, giving total work proportional to $mnD$.

Finally, every landmark belongs to the original dataset, so at most $n$ landmarks can be selected. Substituting the bound $m\leq n$ gives an upper bound proportional to $n^2D$.
\end{proof}

A flat FAISS index changes the implementation constants through vectorization, blocking, optimized kernels, and parallelism, but it does not change this exhaustive asymptotic count. Its advantage is therefore primarily hardware efficiency. By contrast, a metric tree can avoid many pointwise evaluations when node bounds exclude large subsets of the data. Nevertheless, the worst-case cost of a tree range query is still $O(n)$, so without geometric assumptions the worst-case Ball Mapper range query cost remains $O(mn)$ distance evaluations. This distinction is important: empirical speedups and empirical scaling exponents quantify observed performance, whereas asymptotic complexity statements require guarantees that hold over a specified class of inputs.

\subsection{Backend Abstraction and Reference Implementation}

The accompanying implementation exposes a common range query backend interface. Each backend reports whether it is intended to be exact or approximate, its metric and numerical data type, its execution device, whether batched queries are supported, and whether exhaustive distance vectors are available. The latter capability is required by farthest point sampling, which needs distances from each selected landmark to every observation; approximate indexes that return only candidates are therefore excluded from farthest point sampling when they cannot provide distances to all observations.

The exact CPU validation uses three independent implementations: a transparent double-precision floating-point brute force oracle, scikit-learn BallTree, and SciPy cKDTree for Euclidean data. FAISS Flat is exhaustive over its stored single-precision floating-point vectors and is used as the optimized dense vector backend, but comparisons to the double-precision floating-point reference retain an explicit comparison of memberships near the radius threshold because single-precision floating-point representation can differ for points whose distances from the query lie very close to $\varepsilon$. Approximate adapters include FAISS IVF and HNSW. The software also provides optional hnswlib and NVIDIA cuVS adapters through the same interface; these adapters are part of the implementation architecture but are not included in the experiments reported in this article.

This design has two purposes. First, it makes exactness testable independently of a particular library. Second, it represents the theoretical range query oracle through a common software interface: new backends can be added without changing landmark selection, graph construction, coloring, or the validation routines.

\section{Stability Under Approximate Range Queries}\label{sec:approximate-range-guarantees}

FAISS is not intrinsically approximate. Whether a FAISS range query is exact or approximate depends on the index used and on its configuration. In the Ball Mapper construction, this distinction is important because the algorithm requires all points lying inside a prescribed ball, not merely a fixed number of nearest neighbors. 

Throughout this section, let $(X,d)$ be a finite metric space and let $\varepsilon>0$ be the Ball Mapper scale parameter. We now formalize the exact and approximate range query oracles introduced in the Introduction.

\begin{definition}
For $q\in X$ and $r>0$, the \emph{exact range set} centered at $q$ with radius $r$ is
\[
R_r(q) := \{x\in X \mid d(q,x)\leq r\}.
\]
The map $R_r:X\to 2^X$ given by $q \mapsto R_r(q)$ is called the \emph{exact range oracle} at radius $r$.
\end{definition}

An exact range oracle returns precisely the points in $X$ contained in the closed ball of radius $r$ centered at the query point. By contrast, an approximate index may return a set that differs from $R_r(q)$.

\begin{definition}
An \emph{approximate range oracle} at radius $r$ is a map $\widetilde R_r:X\to 2^X$ given by $q \mapsto \widetilde R_r(q)$, where $\widetilde R_r(q)$ is the set returned by an approximate range query procedure.
\end{definition}

The purpose of this section is to describe how differences between the exact range set $R_r(q)$ and the approximate range set $\widetilde R_r(q)$ affect the Ball Mapper cover and graph.

\subsection{Cover Approximation}

We first consider an additive error model. This model assumes that the approximate query is exact except possibly for points whose distance from the query point lies close to the requested radius.

\begin{definition}
Let $0 \leq \delta <r$. An approximate range oracle $\widetilde R_r$ is said to be \emph{$\delta$-accurate} at radius $r$ if, for every $q\in X$,
\[
R_{r-\delta}(q) \subseteq \widetilde R_r(q) \subseteq R_{r+\delta}(q).
\]
\end{definition}

Equivalently, every point at distance at most $r-\delta$ is returned by the approximate query, and no point at distance greater than $r+\delta$ is returned. Only points in the boundary band $r-\delta<d(q,x)\leq r+\delta$ can be ambiguous. This shows that a $\delta$-accurate oracle can make errors only near the boundary of the query ball.

Because $X$ is finite, this boundary description yields an exact robustness criterion whenever the chosen scale is separated from all observed pairwise distances. Define the global radius margin
\[
\gamma_\varepsilon:=\min_{q,x\in X}|d(q,x)-\varepsilon|.
\]
If no observed distance equals $\varepsilon$, then $\gamma_\varepsilon>0$.

\begin{proposition}\label{prop:margin-exact}
If a range oracle is $\delta$-accurate at radius $\varepsilon$ and $0\leq\delta<\gamma_\varepsilon$, then, for every $q\in X$,
\[
\widetilde R_\varepsilon(q)=R_\varepsilon(q).
\]
Consequently, with fixed data order and tie-breaking, the approximate and exact greedy Ball Mapper constructions are identical.
\end{proposition}

\begin{proof}
The inequality $\delta<\gamma_\varepsilon$ implies that no point has distance from $q$ in the closed boundary band $[\varepsilon-\delta,\varepsilon+\delta]$. Hence, \[
R_{\varepsilon-\delta}(q)=R_\varepsilon(q)=R_{\varepsilon+\delta}(q).
\]
The defining inclusions of $\delta$-accuracy force $\widetilde R_\varepsilon(q)$ to equal this common set. The final statement follows from Theorem~\ref{thm:exact-oracle-invariance}.
\end{proof}

Thus, approximate numerical search can still recover the exact Ball Mapper whenever its uncertainty band is smaller than the empirical margin around $\varepsilon$.

\begin{definition}
Let $\rho>0$. A subset $L \subseteq X$ is a \emph{$\rho$-cover} of $X$ if, for every $x\in X$, there exists $l \in L$ such that
\[
d(x,l)\leq\rho.
\]
For $\alpha>0$, the set $L$ is said to be \emph{$\alpha$-separated} if, for every pair of distinct points $l,l'\in L$,
\[
d(l,l')>\alpha.
\]
\end{definition}

An $\varepsilon$-net is simultaneously an $\varepsilon$-cover and an $\varepsilon$-separated set.

\begin{theorem}
Suppose that the greedy Ball Mapper landmark construction is run using a $\delta$-accurate approximate range oracle at radius $\varepsilon$, where $0 \leq \delta < \varepsilon$. Let $L \subseteq X$ be the computed landmark set. Then the following are satisfied:
\begin{enumerate}
\item $L$ is an $(\varepsilon+\delta)$-cover of $X$.
\item $L$ is $(\varepsilon-\delta)$-separated.
\end{enumerate}
\end{theorem}

\begin{proof}
Every point of $X$ is either selected as a landmark or marked as covered by a previous landmark. If $x$ is marked as covered by a landmark $l$, then
\[
x \in \widetilde R_\varepsilon(l).
\]
Since the oracle is $\delta$-accurate,
\[
\widetilde R_\varepsilon(l)\subseteq R_{\varepsilon+\delta}(l).
\]
Hence, $d(x,l) \leq \varepsilon+\delta$. If $x$ is itself selected as a landmark, then $d(x,x)=0 \leq \varepsilon + \delta$. Therefore, $L$ is an $(\varepsilon+\delta)$-cover.

For separation, let $l,l'\in L$ be distinct landmarks, and suppose that $l$ was selected before $l'$. Since $l'$ was uncovered when it was selected, $l' \notin \widetilde R_\varepsilon(l)$. Using the inclusion $R_{\varepsilon-\delta}(l)\subseteq \widetilde R_\varepsilon(l)$,
we obtain
\[
l'\notin R_{\varepsilon-\delta}(l).
\]
Therefore, $d(l,l') > \varepsilon-\delta$. Consequently, $L$ is $(\varepsilon-\delta)$-separated.
\end{proof}

The theorem shows that the output of the approximate greedy construction need not be an exact $\varepsilon$-net. However, if $\delta$ is small relative to $\varepsilon$, then the covering and separation scales are perturbed by at most $\delta$.

The same theorem also controls how many landmarks an approximate construction can produce through standard metric packing and covering numbers. Let $N_X(r)$ be the minimum cardinality of an $r$-cover of $X$, and let $P_X(r)$ be the maximum cardinality of an $r$-separated subset of $X$.

\begin{corollary}\label{cor:packing-covering}
Under the assumptions of the preceding theorem, the approximate landmark set $L$ satisfies
\[
N_X(\varepsilon+\delta)\leq |L|\leq P_X(\varepsilon-\delta).
\]
For exact greedy Ball Mapper, this reduces to
\[
N_X(\varepsilon)\leq |L|\leq P_X(\varepsilon).
\]
\end{corollary}

\begin{proof}
Every $(\varepsilon+\delta)$-cover has at least $N_X(\varepsilon+\delta)$ points, and every $(\varepsilon-\delta)$-separated set has at most $P_X(\varepsilon-\delta)$ points. Applying the two conclusions of the preceding theorem proves the result.
\end{proof}

This identifies one mechanism by which false negatives can increase the number of selected landmarks: weakening separation from $\varepsilon$ to $\varepsilon-\delta$ enlarges the admissible packing number.

\subsection{Nerve Approximation}

We next study how approximate range sets affect the Ball Mapper graph when the landmark set is fixed.

\begin{definition}
Let $L\subseteq X$ be a finite landmark set. For $r>0$, the \emph{exact nerve at scale $r$} is the abstract simplicial complex
\[
\mathcal N_r(L) := \left\{\sigma\subseteq L : \sigma \neq \varnothing,\ \bigcap_{l\in\sigma}R_r(l) \neq \varnothing \right\}.
\]
The corresponding Ball Mapper graph is the $1$-skeleton of $\mathcal N_r(L)$ and is denoted by $G_r(L)$.
\end{definition}

When approximate range sets are used, we obtain an approximate nerve. In the stability results below, $0\leq\delta<\varepsilon$, so $l\in\widetilde R_\varepsilon(l)$ for every landmark $l$ and the returned sets define a simplicial complex on the fixed landmark set.

\begin{definition}
Let $\widetilde R_\varepsilon$ be an approximate range oracle and let $L\subseteq X$ be fixed. The \emph{approximate nerve} at scale $\varepsilon$ is
\[
\widetilde{\mathcal N}_{\varepsilon}(L) := \left\{\sigma\subseteq L : \sigma \neq \varnothing,\ \bigcap_{l\in\sigma}\widetilde R_\varepsilon(l) \neq \varnothing \right\}.
\]
Its $1$-skeleton is denoted by $\widetilde G_\varepsilon(L)$.
\end{definition}

\begin{proposition}\label{prop:nerve_incl}
Suppose that $\widetilde R_\varepsilon$ is $\delta$-accurate at radius $\varepsilon$. For any fixed landmark set $L\subseteq X$,
\[
\mathcal N_{\varepsilon-\delta}(L) \subseteq \widetilde{\mathcal N}_{\varepsilon}(L) \subseteq \mathcal N_{\varepsilon+\delta}(L).
\]
Consequently, $G_{\varepsilon-\delta}(L) \subseteq \widetilde G_\varepsilon(L) \subseteq G_{\varepsilon+\delta}(L)$.
\end{proposition}

\begin{proof}
For every $l\in L$, by assumption, we have
\[
R_{\varepsilon-\delta}(l) \subseteq \widetilde R_\varepsilon(l) \subseteq R_{\varepsilon+\delta}(l).
\]
Therefore, for every nonempty finite subset $\sigma\subseteq L$,
\[
\bigcap_{l\in\sigma}R_{\varepsilon-\delta}(l) \subseteq \bigcap_{l\in\sigma}\widetilde R_\varepsilon(l) \subseteq \bigcap_{l\in\sigma}R_{\varepsilon+\delta}(l).
\]
If the left-hand intersection is nonempty, then the middle intersection is nonempty. If the middle intersection is nonempty, then the right-hand intersection is nonempty. This proves the nerve inclusions. Taking $1$-skeleta gives the graph inclusions.
\end{proof}

The approximate graph at scale $\varepsilon$ is not arbitrary. It lies between two exact Ball Mapper graphs computed at the nearby scales $\varepsilon-\delta$ and $\varepsilon +\delta$. This can also be expressed using the edges.

\begin{definition}
For two landmarks $l_i,l_j\in L$, the \emph{witness radius} is given by
\[
\tau_{ij} := \min_{x\in X} \max\{d(x,l_i),d(x,l_j)\}.
\]
\end{definition}

The witness radius $\tau_{ij}$ is the smallest radius at which some data point lies simultaneously in the two balls centered at $l_i$ and $l_j$. Under the closed ball convention, the edge $(l_i,l_j)$ belongs to $G_r(L)$ precisely when
\[
\tau_{ij} \leq r.
\]

As a consequence of Proposition~\ref{prop:nerve_incl}, the following statements hold:
\begin{enumerate}
\item If $\tau_{ij} \leq \varepsilon-\delta$, then $(l_i,l_j)$ is guaranteed to appear in $\widetilde G_\varepsilon(L)$.
\item If $\tau_{ij} > \varepsilon+\delta$, then $(l_i,l_j)$ is guaranteed not to appear in $\widetilde G_\varepsilon(L)$.
\item An edge between landmarks $l_i$ and $l_j$ satisfying $\varepsilon-\delta<\tau_{ij}\leq\varepsilon+\delta$ may or may not appear in the approximate graph.
\end{enumerate}

Uncertain edges are precisely those whose witnesses occur near the chosen scale parameter.

The pairwise witness radius extends directly to higher-order simplices. For any nonempty $\sigma\subseteq L$, define
\[
\tau_\sigma:=\min_{x\in X}\max_{l\in\sigma}d(x,l).
\]
Then $\sigma\in\mathcal N_r(L)$ exactly when $\tau_\sigma\leq r$. Proposition~\ref{prop:nerve_incl} therefore has the following equivalent threshold interpretation. If $\tau_\sigma\leq\varepsilon-\delta$, then
\[
\sigma\in\widetilde{\mathcal N}_\varepsilon(L).
\]
If $\tau_\sigma>\varepsilon+\delta$, then
\[
\sigma\notin\widetilde{\mathcal N}_\varepsilon(L).
\]
Hence, the uncertainty band applies to the full nerve, not only to graph edges.

For some approximate methods, an additive range error is not the most natural model. Instead, the computed distance may approximate the true distance multiplicatively.

\begin{definition}
Let $0\leq\eta<1$. An approximate distance function $\widetilde d: X \times X \to \mathbb R$ is said to have \emph{relative error at most $\eta$} if
\[
(1-\eta)d(q,x) \leq \widetilde d(q,x) \leq (1+\eta)d(q,x)
\]
for all $q,x \in X$.
\end{definition}

Given such an approximate distance, the approximate range set is
\[
\widetilde R_\varepsilon(q) := \{x\in X \mid \widetilde d(q,x) \leq \varepsilon\}.
\]

\begin{proposition}\label{prop:multiplicative-range}
If $\widetilde d$ has relative error at most $\eta$, then for every $q\in X$,
\[
R_{\varepsilon/(1+\eta)}(q) \subseteq \widetilde R_\varepsilon(q) \subseteq R_{\varepsilon/(1-\eta)}(q).
\]
\end{proposition}

\begin{proof}
Suppose first that $x\in R_{\varepsilon/(1+\eta)}(q)$. Then
\[
d(q,x)\leq\frac{\varepsilon}{1+\eta}.
\]
The upper relative error bound gives $\widetilde d(q,x)\leq(1+\eta)d(q,x)\leq\varepsilon$. Hence, $x\in\widetilde R_\varepsilon(q)$.

Conversely, suppose that $x\in\widetilde R_\varepsilon(q)$. Then $\widetilde d(q,x)\leq\varepsilon$. The lower relative error bound gives $(1-\eta)d(q,x)\leq\widetilde d(q,x)\leq\varepsilon$. Therefore,
\[
d(q,x)\leq\frac{\varepsilon}{1-\eta},
\]
so $x\in R_{\varepsilon/(1-\eta)}(q)$.
\end{proof}

Therefore, the preceding cover, separation, nerve, and graph results also apply in the multiplicative setting. The inner radius is $\varepsilon/(1+\eta)$, and the outer radius is $\varepsilon/(1-\eta)$.

\subsection{Range Query Errors}

The approximate range set may differ from the exact range set in two basic ways.

\begin{definition}
Let $q\in X$. The set of \emph{false negatives} at radius $\varepsilon$ is
\[
\operatorname{FN}_\varepsilon(q) := R_\varepsilon(q) \setminus \widetilde R_\varepsilon(q).
\]
The set of \emph{false positives} at radius $\varepsilon$ is
\[
\operatorname{FP}_\varepsilon(q) := \widetilde R_\varepsilon(q)\setminus R_\varepsilon(q).
\]
\end{definition}

False negatives are true members of the exact ball that are not returned by the approximate query. False positives are points returned by the approximate query even though they do not belong to the exact ball.

The two error types affect Ball Mapper differently. During landmark selection, a false negative may leave a point uncovered even though it lies inside an existing landmark ball. Such a point may later be selected as an additional landmark. A false positive may mark a point as covered even though it does not lie inside the exact ball. This can destroy exact $\varepsilon$-coverage.

For a fixed landmark set, false negatives may remove genuine intersection witnesses and hence delete edges. False positives may introduce spurious witnesses and hence add edges.

\begin{definition}
For $q\in X$, define the \emph{range recall} and \emph{range precision} by
\[
\operatorname{recall}_\varepsilon(q) := \frac{|R_\varepsilon(q) \cap \widetilde R_\varepsilon(q)|}{|R_\varepsilon(q)|},
\]
and
\[
\operatorname{precision}_\varepsilon(q) := \frac{|R_\varepsilon(q) \cap \widetilde R_\varepsilon(q)|}{|\widetilde R_\varepsilon(q)|},
\]
whenever the denominators are non-zero.
\end{definition}

Range recall measures the proportion of true ball members returned by the approximate query. Range precision measures the proportion of returned points that are true ball members.

A common way to eliminate false positives is to use an approximate index only to generate a candidate set $C_\varepsilon(q)\subseteq X$, and then evaluate the exact distance $d(q,x)$ for each candidate $x \in C_\varepsilon(q)$. The \emph{verified range set} is defined by
\[
\widetilde R_\varepsilon(q) := \{x\in C_\varepsilon(q) \mid d(q,x)\leq\varepsilon\}.
\]
Since membership is decided using the exact distance, we have
\[
\widetilde R_\varepsilon(q)\subseteq R_\varepsilon(q).
\]
Thus, exact verification eliminates false positives. False negatives may remain if the candidate set $C_\varepsilon(q)$ omits points belonging to $R_\varepsilon(q)$.

\begin{definition}
An approximate range oracle is called \emph{conservative} at radius $\varepsilon$ if
\[
\widetilde R_\varepsilon(q)\subseteq R_\varepsilon(q)
\]
for every $q\in X$.
\end{definition}

\begin{proposition}\label{prop:graph_incl}
Suppose that the greedy Ball Mapper construction uses a conservative approximate range oracle at radius $\varepsilon$. Let $L \subseteq X$ be the resulting landmark set. Then the following statements hold:
\begin{enumerate}
\item The exact balls of radius $\varepsilon$ centered at points of $L$ cover $X$.
\item For a fixed landmark set $L$, $\widetilde G_\varepsilon(L)\subseteq G_\varepsilon(L)$.
\end{enumerate}
\end{proposition}

\begin{proof}
If $x$ is marked as covered by a landmark $l$, then $x\in \widetilde R_\varepsilon(l)$. Since the oracle is conservative, $x \in R_\varepsilon(l)$ or $d(x,l) \leq \varepsilon$. If $x$ is itself selected as a landmark, then $d(x,x) = 0 \leq \varepsilon$. Therefore, the exact $\varepsilon$-balls centered at landmarks in $L$ cover $X$.

Now, suppose $(l_i, l_j)$ is an edge of $\widetilde G_\varepsilon(L)$. Then there exists $x \in X$ such that
\[
x\in \widetilde R_\varepsilon(l_i)\cap \widetilde R_\varepsilon(l_j).
\]
Since the oracle is conservative,
\[
x\in R_\varepsilon(l_i)\cap R_\varepsilon(l_j).
\]
Thus, $(l_i, l_j)$ is also an edge of $G_\varepsilon(L)$.
\end{proof}

The resulting landmark set $L$ need not be $\varepsilon$-separated. Separation may fail because a point $x$ satisfying $d(x,l) \leq \varepsilon$ may be omitted from $\widetilde R_\varepsilon(l)$ by a false negative. Then $x$ remains uncovered and may later be selected as another landmark. Hence, two distinct landmarks may lie at distance at most $\varepsilon$.

The graph inclusion in Proposition~\ref{prop:graph_incl} is a statement for a fixed landmark set. If approximate range queries are also used during landmark selection, then the exact and approximate constructions may produce different landmark sets, and the resulting graphs need not have the same vertex set.

\subsubsection{Witness Redundancy and Edge Survival}
For a fixed landmark set and conservative approximate ranges, graph stability is governed not only by range recall but also by the number of distinct data points witnessing each overlap. Define the exact witness set by
\[
W_{ij}:=R_\varepsilon(l_i)\cap R_\varepsilon(l_j).
\]
Let $m_{ij}:=|W_{ij}|$ denote its cardinality, and let $F_i:=R_\varepsilon(l_i)\setminus\widetilde R_\varepsilon(l_i)$ be the false negative set for landmark $l_i$.

\begin{proposition}\label{prop:witness-redundancy}
Suppose the approximate oracle is conservative and the landmark set is fixed. An exact edge $(l_i,l_j)$ survives in the approximate graph if and only if
\[
W_{ij}\setminus(F_i\cup F_j)\neq\varnothing.
\]
In particular, the sufficient condition
\[
m_{ij}>|F_i\cap W_{ij}|+|F_j\cap W_{ij}|
\]
guarantees that the edge survives. A still simpler sufficient condition is $m_{ij}>|F_i|+|F_j|$.
\end{proposition}

\begin{proof}
Under conservative search, the approximate common witnesses are exactly
\[
\widetilde R_\varepsilon(l_i)\cap\widetilde R_\varepsilon(l_j)
= W_{ij}\setminus(F_i\cup F_j).
\]
The edge survives precisely when this set is nonempty. If every exact witness were removed, then $W_{ij}\subseteq F_i\cup F_j$, which would imply
$|W_{ij}|\leq |F_i\cap W_{ij}|+|F_j\cap W_{ij}|$.
The contrapositive gives the sufficient condition.
\end{proof}

Thus, two edges with the same geometric scale margin can have very different robustness: an overlap supported by many data points can tolerate several missed memberships, whereas an edge with a single witness disappears as soon as that witness is missed by either endpoint. The same argument extends to a simplex $\sigma$: it survives whenever at least one exact common witness is retained by every landmark in $\sigma$. This gives a direct theoretical explanation for the empirical observation that clustered or low-dimensional structured datasets can preserve graph connectivity despite imperfect range recall.

\subsubsection{Stability of Average Colorings Under Range Errors}
Approximate range queries can change a Ball Mapper visualization even when the landmark set is held fixed, because the color of a vertex is computed from its returned membership set. This effect can also be bounded directly.

Let $X_i=R_\varepsilon(l_i)$ and $\widetilde X_i=\widetilde R_\varepsilon(l_i)$, and assume that both sets are nonempty. Define the exact mean color by
\[
c_i=\frac{1}{|X_i|}\sum_{x\in X_i}f(x),
\]
and define the approximate mean color by
\[
\widetilde c_i=\frac{1}{|\widetilde X_i|}\sum_{x\in\widetilde X_i}f(x).
\]
Write $r_i$ and $p_i$ for the range recall and precision at $l_i$, and define the local oscillation
\[
\operatorname{osc}_i(f)=\max_{x\in X_i\cup\widetilde X_i}f(x)-\min_{x\in X_i\cup\widetilde X_i}f(x).
\]

\begin{theorem}\label{thm:color-range-error}
For every fixed landmark $l_i$,
\[
|\widetilde c_i-c_i|
\leq \operatorname{osc}_i(f)\bigl(1-\min\{r_i,p_i\}\bigr).
\]
If the oracle is $\delta$-accurate and $f$ is $k$-Lipschitz, then
\[
|\widetilde c_i-c_i|
\leq2k(\varepsilon+\delta)\bigl(1-\min\{r_i,p_i\}\bigr).
\]
For a conservative oracle, $p_i=1$ and $X_i\cup\widetilde X_i=X_i$, so the sharper specialization is
\[
|\widetilde c_i-c_i|\leq2k\varepsilon(1-r_i).
\]
\end{theorem}

\begin{proof}
Let $U_i$ and $\widetilde U_i$ denote the uniform probability measures on $X_i$ and $\widetilde X_i$. For two finite uniform measures,
\[
\|U_i-\widetilde U_i\|_{\mathrm{TV}}
=1-\frac{|X_i\cap\widetilde X_i|}{\max\{|X_i|,|\widetilde X_i|\}}
=1-\min\{r_i,p_i\}.
\]
The difference between expectations of a function whose range has width $\operatorname{osc}_i(f)$ is bounded by that width times total variation, proving the first inequality. Under $\delta$-accuracy, both membership sets lie in $R_{\varepsilon+\delta}(l_i)$, whose diameter is at most $2(\varepsilon+\delta)$; Lipschitz continuity therefore gives $\operatorname{osc}_i(f)\leq2k(\varepsilon+\delta)$. Under conservative search, $\widetilde X_i\subseteq X_i$, so precision equals one and the union is contained in the exact $\varepsilon$-ball.
\end{proof}

This separates two sources of visual variation: geometric variation of $f$ over the cover, controlled by the coloring results in Section~2.3, and computational variation caused by imperfect range membership, controlled here by recall and precision.

\subsection{Approximation Methods in FAISS}

FAISS contains several indexing methods \parencite{faissIndexesWiki}. Their effect on Ball Mapper can be interpreted through the range set framework introduced above.

A flat index, such as \texttt{IndexFlatL2}, compares the query with every stored vector \parencite{faissIndexFlatDocs}. Hence, it computes the exact range set, up to some floating-point precision near the boundary.

First, consider a \emph{non-exhaustive index}. Such an index examines only a subset of the data points during a query. Let $C(q)\subseteq X$ denote the set of candidate points examined for the query point $q$. This is a conservative approximation. Since membership is checked using the exact distance $d$, no point outside the exact ball can be returned. Hence, false positives are impossible. However, false negatives may occur.

Examples of this type include inverted file indexes such as \texttt{IndexIVFFlat}, where the dataset is divided into cells and only selected cells are searched, and graph-based indexes such as \texttt{IndexHNSWFlat}, where the query explores only part of a proximity graph \parencite{Malkov2020HNSW}.

Second, consider indexes based on compressed distance computation. In this case, a vector $x$ is replaced by a compressed representation, and the query uses an approximate distance $\widetilde d(q,x)$ instead of the exact distance $d(q,x)$. Because $\widetilde d(q,x)$ may be smaller or larger than $d(q,x)$, the set $\widetilde R_\varepsilon(q)$ need not be contained in $R_\varepsilon(q)$, nor does it necessarily contain $R_\varepsilon(q)$. Thus, both false positives and false negatives may occur.

Examples of compressed distance indexes include scalar quantization and product quantization \parencite{Jegou2011PQ} indexes, such as \texttt{IndexSQ} and \texttt{IndexPQ}.

Finally, some FAISS indexes combine both mechanisms. For example, an inverted file index with product quantization first selects a candidate set $C(q)$ and then evaluates approximate distances using compressed vector representations \parencite{SivicZisserman2003}. False negatives and false positives, as explained above, may occur.

The main cases can be summarized as follows, following the FAISS index taxonomy \parencite{faissIndexesWiki}:
\begin{center}
\begin{tabular}{lll}
\toprule
\textbf{Index type} &
\textbf{Source of approximation} &
\textbf{Possible range errors} \\
\midrule
\texttt{IndexFlatL2} &
Exact &
None \\
\texttt{IndexIVFFlat} &
Only selected cells are examined &
False negatives \\
\texttt{IndexHNSWFlat} &
Partial graph traversal &
False negatives \\
\texttt{IndexPQ} or \texttt{IndexSQ} &
Compressed distance evaluation &
False positives and false negatives \\
\texttt{IndexIVFPQ} &
Cell selection and compression &
False positives and false negatives \\
\bottomrule
\end{tabular}
\end{center}

Thus, the deterministic guarantees proved above should be interpreted conditionally. If an approximate FAISS configuration satisfies an additive or multiplicative range set error bound, then the corresponding Ball Mapper cover and graph satisfy the stated guarantees. If no such bound is available, then the approximation should instead be evaluated empirically by comparing it with the exhaustive \texttt{IndexFlatL2} baseline, using quantities such as range recall and range precision.

\section{Exact Backend Benchmark}

We first evaluate exact CPU implementations of the greedy Ball Mapper cover construction using BallTree and exhaustive FAISS Flat, with \emph{pyBallMapper} (pyBM) as an external software baseline. This benchmark is distinct from the approximation experiments in Section~\ref{sec:approx-experiments}: it asks whether the same exact range sets can be realized more efficiently, not how much structural error can be tolerated.

All experiments were run on a machine with a 13th-generation Intel Core i5-13420H processor (2.10 GHz) featuring 8 cores and 12 threads, and 16 GB of RAM. Computations were carried out on the CPU using integrated Intel UHD Graphics; no GPU acceleration was used. All methods were implemented in Python 3.12 with scikit-learn 1.7, FAISS 1.13, and pyBallMapper 0.3 libraries.

Datasets were generated to investigate the effects of dataset size and dimension. Across all experiments, the dataset size was set to $n \in \left\{ 100,500,1000,2000,5000 \right\}$, and the dimension to $D \in \left\{ 10,50,100,200,500,1000 \right\}$, with 65 repeated trials conducted for each $(n,D)$ combination.

In all experiments, the Euclidean distance is used for the $\varepsilon$-net construction. The leaf size of the Ball Tree data structure is set to 40, and FAISS is configured to allow up to 12 CPU threads. The main Python library is available through a GitHub \href{https://github.com/jhnrckmnznrs/fast-ballmapper}{repository}.

\subsection{Runtime and Speedup}

We investigate the runtime behavior and relative speedup of Ball Tree, FAISS, and pyBallMapper by examining how performance scales with problem size. We conduct a two-stage benchmarking procedure, first varying the number of data points $n$ at fixed dimensionality, and then varying the dimensionality $D$ at fixed dataset size.

\begin{figure}[ht]
\centering
\includegraphics[width=\textwidth]{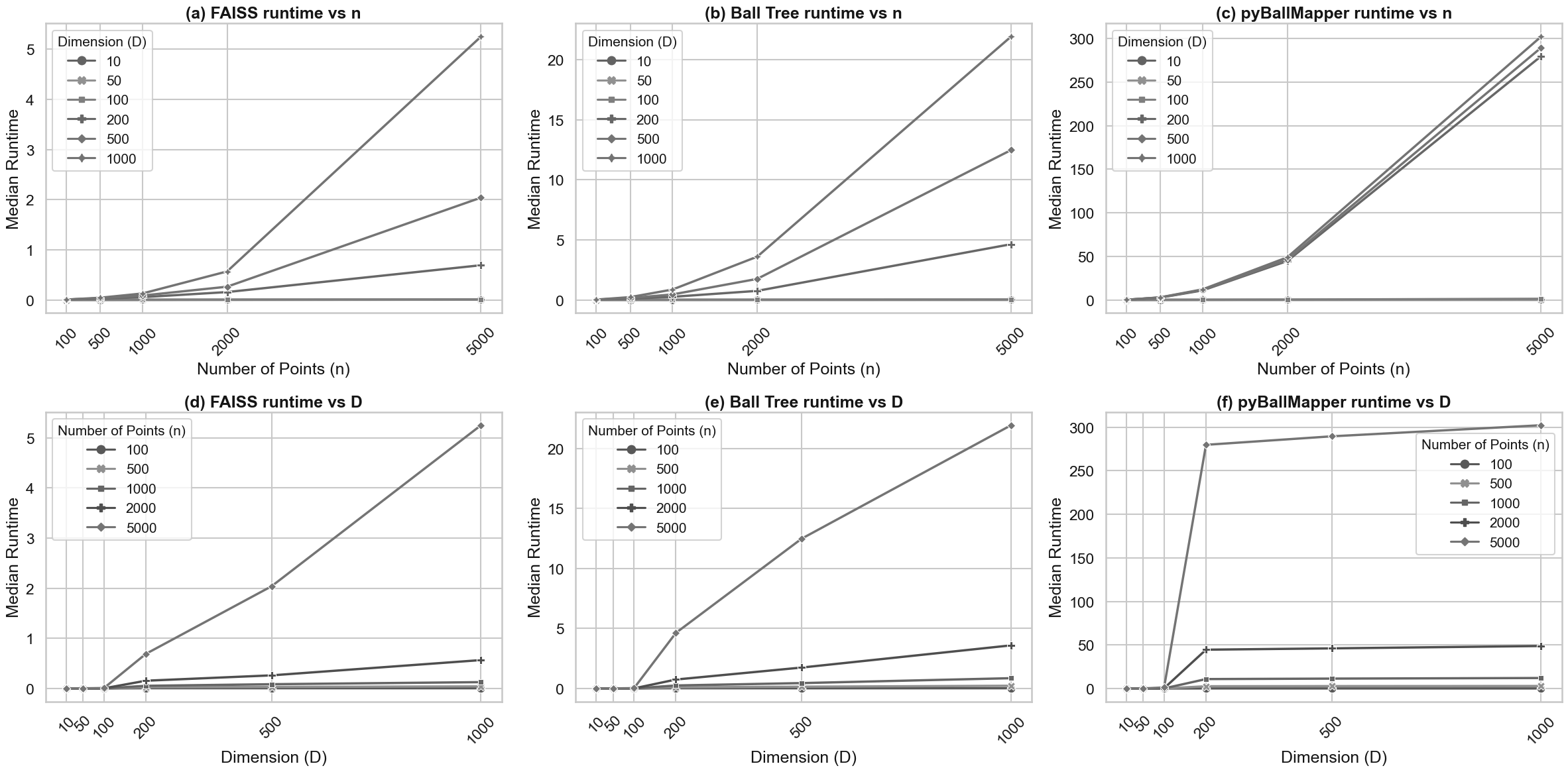}
\caption{Median runtime across varying dataset sizes $n$ and dimensions $D$. Panels (a)--(c) show runtime as a function of $n$; panels (d)--(f) show runtime as a function of $D$. Series are distinguished by marker shape and grayscale intensity.}
\label{fig:runtime}
\end{figure}

Figure~\ref{fig:runtime} reports the median runtime of the algorithms across the experimental conditions described above. The runtime of all methods increases with the number of data points $n$, though with markedly different scaling behavior. FAISS consistently exhibits the best runtime performance, maintaining low runtimes even as $n$ becomes large. Ball Tree scales reasonably with $n$ at fixed dimensionality, but shows higher runtimes than FAISS across all dataset sizes. In contrast, pyBallMapper displays a much steeper increase in runtime as $n$ grows, becoming substantially more expensive for large datasets.

Figure~\ref{fig:runtime} illustrates the effect of increasing dimensionality on runtime. FAISS remains comparatively efficient even at high $D$, demonstrating strong robustness to increasing dimensionality. Ball Tree performs well at low dimensions but degrades rapidly as $D$ increases, reflecting the impact of the curse of dimensionality on tree-based nearest-neighbor methods. pyBallMapper shows the most pronounced sensitivity to dimensionality, with runtimes increasing sharply and becoming especially costly for $D \geq 200$.

While runtime provides insight into raw performance, it does not directly convey the relative computational advantage of alternative methods over a baseline. 
To better quantify how much faster FAISS and Ball Tree are compared with pyBallMapper under identical experimental conditions, we report the \emph{speedup} of a method relative to pyBM as
\[
S=\frac{T_{\mathrm{pyBM}}}{T_{\mathrm{method}}}.
\]
Thus, $S>1$ means that the method is faster than pyBM, and $S=100$ corresponds to a hundredfold reduction in runtime. This relative measure complements the absolute runtime analysis.

\begin{figure}[ht]
\centering
\includegraphics[width=0.85\textwidth]{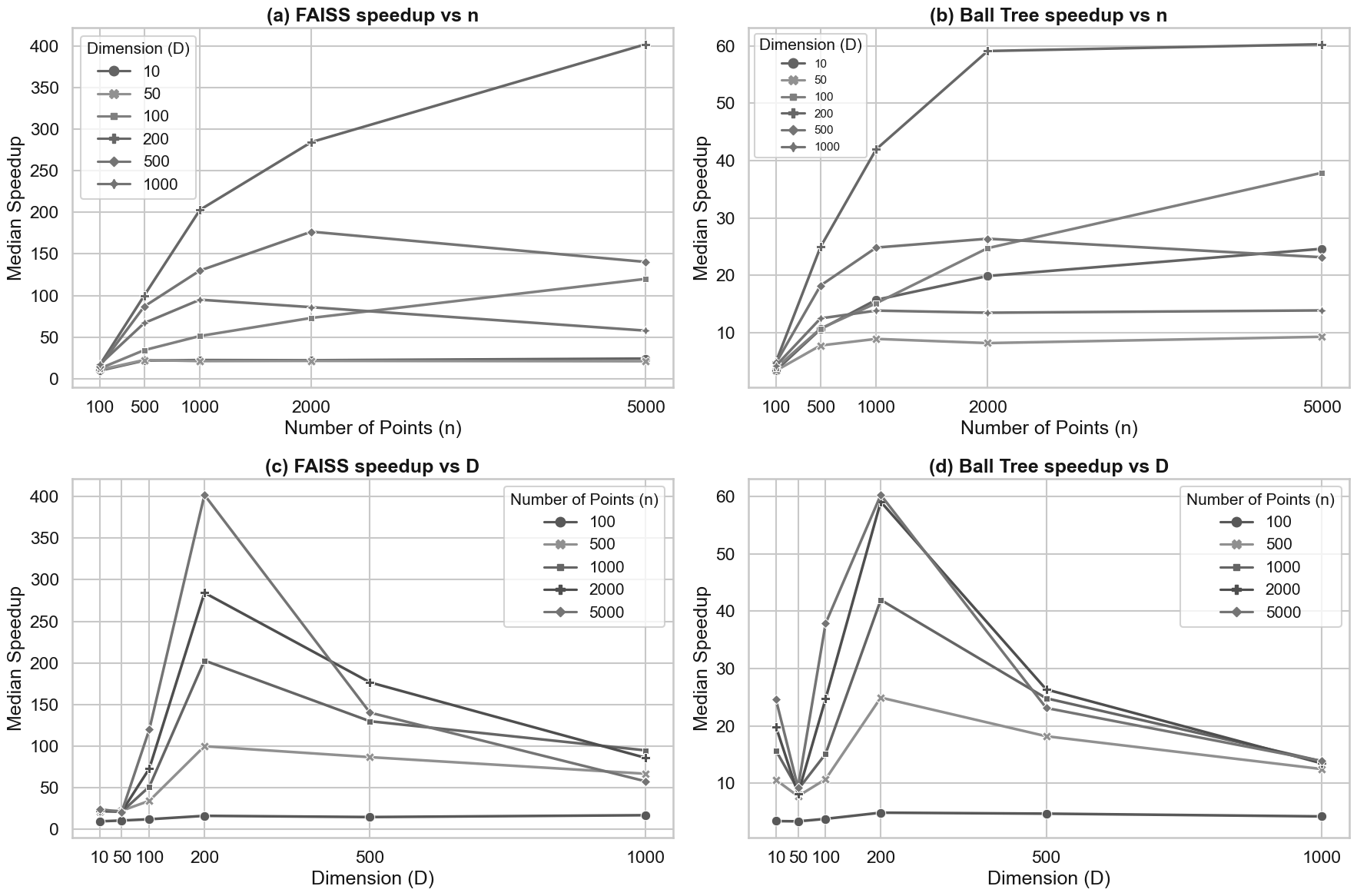}
\caption{Median speedup of FAISS and Ball Tree relative to pyBallMapper across varying dataset sizes $n$ and dimensions $D$. Series are distinguished by marker shape and grayscale intensity.}
\label{fig:speedup}
\end{figure}

Figure~\ref{fig:speedup} shows the median speedup of FAISS and ball tree relative to pyBallMapper as the number of data points $n$ increases. For both methods, speedup grows with $n$, indicating that the computational cost of pyBallMapper scales more rapidly with dataset size than that of the proposed approaches. FAISS exhibits the largest gains, achieving speedups exceeding two orders of magnitude for large datasets, while Ball Tree shows consistent but more modest improvements across the range of $n$.

As illustrated in Figure~\ref{fig:speedup}, speedup also increases with dimensionality $D$. This trend reflects the sharp degradation of pyBallMapper performance in high-dimensional settings. FAISS again achieves the most substantial improvements, reaching speedups of over two orders of magnitude at moderate to high dimensions (approximately $200$). Ball Tree benefits from increased dimensionality as well, but to a lesser extent, consistent with its sensitivity to the curse of dimensionality.

Within these synthetic CPU benchmarks, FAISS gives the lowest runtime across the tested large and high-dimensional conditions, although the magnitude of the advantage depends on hardware and thread count. This robustness is reflected in both low absolute runtimes and large speedups relative to pyBallMapper. Ball Tree is well suited for low-dimensional settings, although its runtime and relative performance are sensitive to the choice of leaf size, which can significantly affect query efficiency. In contrast, the steep runtime growth of pyBallMapper limits its practicality beyond small or low-dimensional regimes.

The corresponding memory measurements and empirical scaling regressions are reported in Appendix~\ref{app:extended-benchmark}. They are kept separate from the main runtime comparison because the fitted exponents describe finite experimental regimes rather than asymptotic complexity.

\section{Approximation Experiments}\label{sec:approx-experiments}

In this section, we evaluate the structural consequences of replacing exhaustive Euclidean range queries by non-exhaustive FAISS queries. The experiments use three independent random seeds for each of three synthetic geometries. The purpose is not to assume that approximate search is faster at the present problem size, but to quantify the fidelity of Ball Mapper memberships, overlap witnesses, graph edges, colors, and landmark sets as search effort is varied.

\subsection{Experimental Setup and Validation}

All experiments use finite point clouds $X=\{x_1,\ldots,x_n\}\subset\mathbb R^d$ with Euclidean distance, with $n=20000$, $d=50$, target ball size $k=50$, and seeds $0$, $1$, and $2$. We consider an isotropic Gaussian point cloud, a Gaussian mixture, and a noisy one-dimensional curve embedded in $\mathbb R^{50}$. These represent unstructured, clustered, and low intrinsic-dimensional geometries.

For each dataset and seed, $\varepsilon$ is selected by the target occupancy rule: for each observation we compute the distance to its $k$th nearest neighbor, counting the observation itself as the first neighbor, and take the median of these distances. Landmark selection for the fixed landmark experiments is performed with exhaustive FAISS \texttt{IndexFlatL2}. The reference cover on that landmark set is then rebuilt independently with the double-precision floating-point brute force oracle, so that exact membership is not defined by the same backend being evaluated. The FAISS Flat cover is also compared with this references.

Approximate fixed landmark covers are constructed with \texttt{IndexIVFFlat}, using
\[
\texttt{nprobe}\in\{1,2,4,8,16,32,64,128,256\},
\]
and with \texttt{IndexHNSWFlat}, using
\[
\texttt{efSearch}\in\{16,32,64,128,256,512\}.
\]
The HNSW maximum number of candidate points is set to $4096$. Across all nine datasets, the largest exact ball contained $1751$ observations, so the maximum number of candidate points exceeds every reference ball size. Thus, a loss of recall cannot be attributed merely to choosing a candidate limit smaller than the true ball, although the approximate candidate search can still omit true members. All experiments use the closed ball convention $d(x,l)\le\varepsilon$, implemented through the common floating-point boundary convention based on $\operatorname{nextafter}$ described in Section~\ref{sectionQuery}.

Before interpreting approximation error, we independently validated the closed ball implementation against the double-precision floating-point brute force reference oracle. The test compares greedy landmark selection, fixed landmark memberships, and the resulting graph under the same $\varepsilon$ and deterministic data order. Table~\ref{tab:exactness-gate} reports two Gaussian cases. In both cases BallTree and SciPy cKDTree selected exactly the same landmark indices as brute force; membership precision, recall, and Jaccard index were one; there were no false positive or false negative memberships; and the edge sets were identical.

\begin{table}[ht]

\centering
\caption{Closed ball exact validation after the unified $\operatorname{nextafter}$ radius policy. BallTree and SciPy cKDTree both agree with the double-precision floating-point brute force oracle on landmarks, memberships, and graph edges.}
\label{tab:exactness-gate}
\begin{tabular}{rrrrrrcc}
\toprule
$n$ & $d$ & $\varepsilon$ & $|L|$ & $|E|$ & Membership Jaccard & Exact landmarks & Exact edges \\
\midrule
1000 & 16 & 3.772222 & 297 & 3686 & 1.000 & yes & yes \\
5000 & 32 & 5.964502 & 1263 & 85588 & 1.000 & yes & yes \\
\bottomrule
\end{tabular}
\end{table}

We use the following evaluation metrics.

In experiments with a fixed landmark set, the exact and approximate constructions share the same landmark set. We report membership recall and precision, edge recall and precision, and witness recall. If
\[
W_{ij}=C_i\cap C_j
\]
is the set of observations witnessing the exact edge $\{i,j\}$, witness recall is the fraction of exact witness incidences retained by the approximate cover. Thus, membership recall measures point to ball fidelity, witness recall measures overlap fidelity, and edge recall measures whether at least one witness remains for each exact edge.

For a scalar field $f$, we also compare mean colors computed from the exact and approximate balls. The experiments use the first coordinate, which is $1$-Lipschitz under the Euclidean metric. The color error bound in Theorem~\ref{thm:color-range-error} is evaluated for every approximate ball.

In experiments where each approximate method selects its own landmarks, each approximate index constructs its own landmark set. We report the relative landmark count, normalized covering radius
\[
\rho(L_{\mathrm{approx}})/\varepsilon,
\]
and normalized minimum landmark separation
\[
s(L_{\mathrm{approx}})/\varepsilon.
\]

\subsection{Reference Geometry and Fixed Landmark Fidelity}

The reference constructions were stable across the three random seeds within each geometry. Table~\ref{tab:reference-geometry} reports mean and standard deviation. The Gaussian point clouds generated many more graph edges than the noisy curve and also exhibited the largest variation in maximum ball occupancy; nevertheless, the fixed maximum number of candidate points of $4096$ exceeded the maximum occupancy in every run.

\begin{table}[ht]

\centering
\caption{Reference Ball Mapper geometry across three seeds, reported as mean $\pm$ standard deviation.}
\label{tab:reference-geometry}
\begin{tabular}{lrrrr}
\toprule
Dataset & $\varepsilon$ & Landmarks & Edges & Maximum ball size \\
\midrule
Gaussian & $7.5471\pm0.0100$ & $6361.3\pm21.5$ & $927825.7\pm8286.3$ & $969.7\pm676.8$ \\
Mixture & $5.6060\pm0.0080$ & $5751.3\pm56.3$ & $505126.7\pm7440.7$ & $451.7\pm30.7$ \\
Curve & $0.32501\pm0.00030$ & $1695.7\pm18.5$ & $6091.7\pm57.2$ & $106.3\pm5.5$ \\
\bottomrule
\end{tabular}
\end{table}

Across all fixed landmark approximate configurations, membership precision, witness precision, and edge precision were equal to one. No approximate run introduced a false positive membership, an added witness, or a spurious edge. The observed approximation was therefore conservative in this experiment: increasing approximation removed memberships and overlap witnesses rather than inventing them. Increasing \texttt{nprobe} or \texttt{efSearch} monotonically improved fidelity and converged to the exact construction.

\begin{figure}[ht]
\centering
\includegraphics[width=0.72\textwidth]{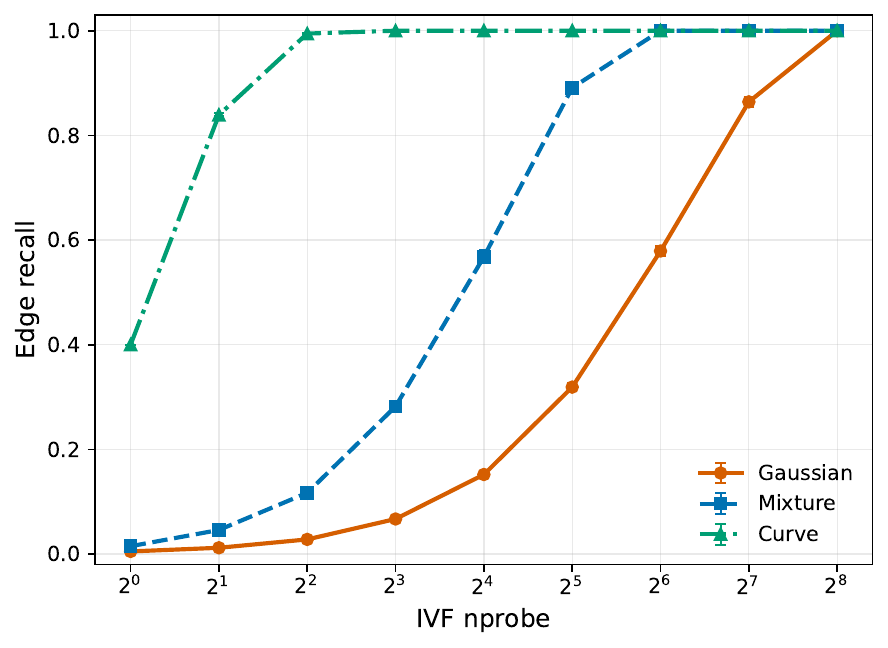}
\caption{Fixed-landmark edge recall for IVF across three seeds. Error bars show one standard deviation. Dataset identity is encoded by both color and marker shape.}
\label{fig:final-edge-ivf}
\end{figure}

\begin{figure}[ht]
\centering
\includegraphics[width=0.72\textwidth]{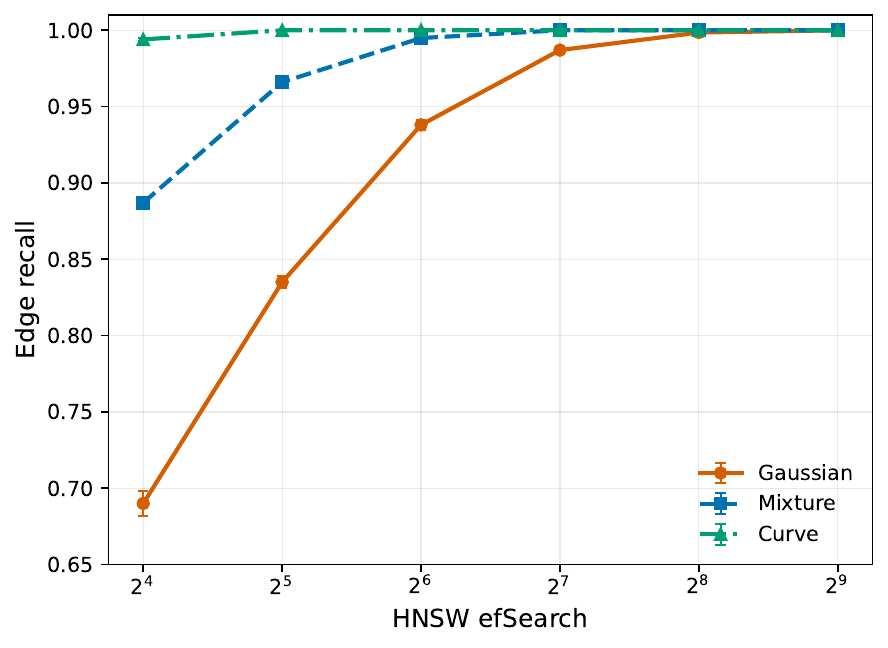}
\caption{Fixed-landmark edge recall for HNSW across three seeds. Error bars show one standard deviation. Dataset identity is encoded by both color and marker shape.}
\label{fig:final-edge-hnsw}
\end{figure}

Figures~\ref{fig:final-edge-ivf} and~\ref{fig:final-edge-hnsw} show mean edge recall with one standard deviation across seeds. The Gaussian geometry was consistently the most sensitive. For Gaussian data, IVF required exhaustive probing (\texttt{nprobe}=256) to exceed $0.99$ edge recall in every seed, whereas HNSW reached mean edge recall of $0.9985\pm0.0001$ at \texttt{efSearch}=256. The mixture required substantially less search effort: IVF was exact at \texttt{nprobe}=64 and HNSW achieved $0.9949\pm0.0005$ for edge recall at \texttt{efSearch}=64. The noisy curve was the most stable: IVF at \texttt{nprobe}=4 already reached $0.9945\pm0.0015$ for edge recall, while HNSW at \texttt{efSearch}=32 recovered every exact edge in all three seeds.

Table~\ref{tab:representative-fidelity} compares the same moderate setting, $64$, across datasets. This common setting makes the geometry effect explicit. In the Gaussian case, HNSW preserved substantially more memberships, witnesses, and edges than IVF. For mixture and curve data, the same search effort was already nearly exact or exact.

\begin{table}[ht]

\centering
\small
\caption{Representative fixed landmark results across three seeds, reported as mean $\pm$ standard deviation. Time/Flat is the cover query time divided by exhaustive FAISS Flat time for the same dataset and seed.}
\label{tab:representative-fidelity}
\resizebox{\textwidth}{!}{%
\begin{tabular}{llccccc}
\toprule
Dataset & Setting & Membership recall & Witness recall & Edge recall & Color MAE & Time/Flat \\
\midrule
Gaussian & IVF, \texttt{nprobe}=64 & $0.760\pm0.004$ & $0.492\pm0.009$ & $0.579\pm0.009$ & $0.0620\pm0.0011$ & $4.96\pm0.20$ \\
Gaussian & HNSW, \texttt{efSearch}=64 & $0.911\pm0.002$ & $0.902\pm0.002$ & $0.938\pm0.003$ & $0.0232\pm0.0011$ & $4.23\pm0.18$ \\
\addlinespace
Mixture & IVF, \texttt{nprobe}=64 & $1.000\pm0.000$ & $1.000\pm0.000$ & $1.000\pm0.000$ & $0.0000\pm0.0000$ & $4.15\pm0.09$ \\
Mixture & HNSW, \texttt{efSearch}=64 & $0.985\pm0.001$ & $0.986\pm0.000$ & $0.995\pm0.000$ & $0.0026\pm0.0002$ & $3.29\pm0.09$ \\
\addlinespace
Curve & IVF, \texttt{nprobe}=64 & $1.000\pm0.000$ & $1.000\pm0.000$ & $1.000\pm0.000$ & $0.0000\pm0.0000$ & $5.10\pm0.35$ \\
Curve & HNSW, \texttt{efSearch}=64 & $1.000\pm0.000$ & $1.000\pm0.000$ & $1.000\pm0.000$ & $0.0000\pm0.0000$ & $2.40\pm0.36$ \\
\bottomrule
\end{tabular}%
}
\end{table}

\subsection{Structural Error Propagation}

\begin{figure}[ht]
\centering
\includegraphics[width=0.72\textwidth]{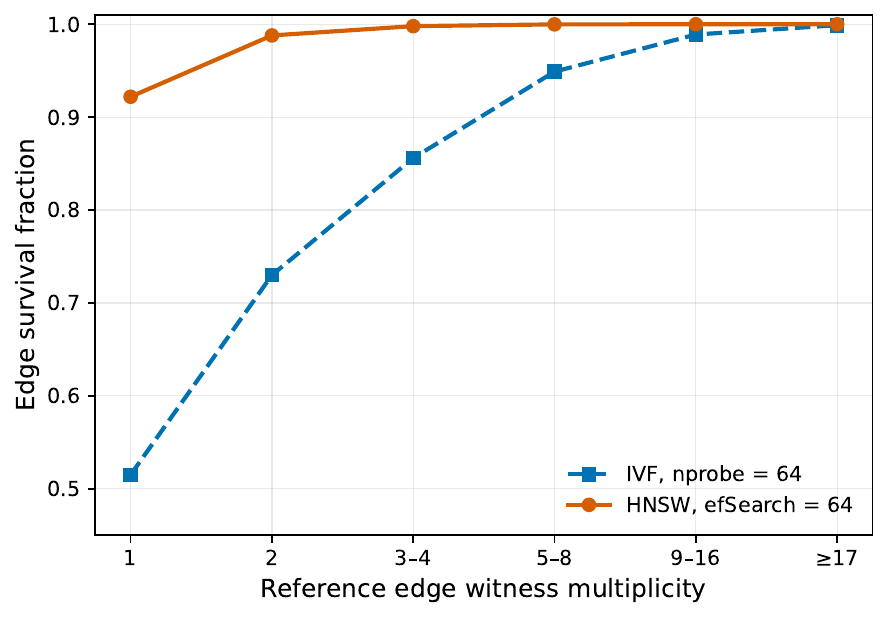}
\caption{Pooled Gaussian edge survival across three seeds as a function of exact witness multiplicity. The two approximate backends are distinguished by both color and line style.}
\label{fig:witness-survival}
\end{figure}

The edge level results provide direct empirical support for the witness redundancy analysis. Figure~\ref{fig:witness-survival} pools the three Gaussian seeds at moderate search effort. Under HNSW with \texttt{efSearch}=64, edges supported by a single exact witness observed a survival proportion of $0.922$, compared with $0.988$ for two witnesses, $0.998$ for three to four witnesses, and greater than $0.9998$ for five to eight witnesses; all edges with at least nine witnesses survived. The same monotone relationship appears under the IVF \texttt{nprobe}=64 approximation, which produced larger membership errors: survival increased from $0.515$ for edges with a single witness to $0.730$ for two witnesses, $0.856$ for three to four, $0.949$ for five to eight, and $0.999$ for at least seventeen witnesses.

This result clarifies why membership and edge recall need not change at the same rate. Redundant witnesses can protect an edge after some of its member observations are missed. Conversely, when approximation is sufficiently severe, many witnesses can be removed simultaneously and edge recall can fall below membership recall. Witness redundancy therefore attenuates moderate membership error; it does not imply robustness to arbitrarily large membership errors.

We next consider the stability of mean vertex colors.

\begin{figure}[ht]
\centering
\includegraphics[width=0.66\textwidth]{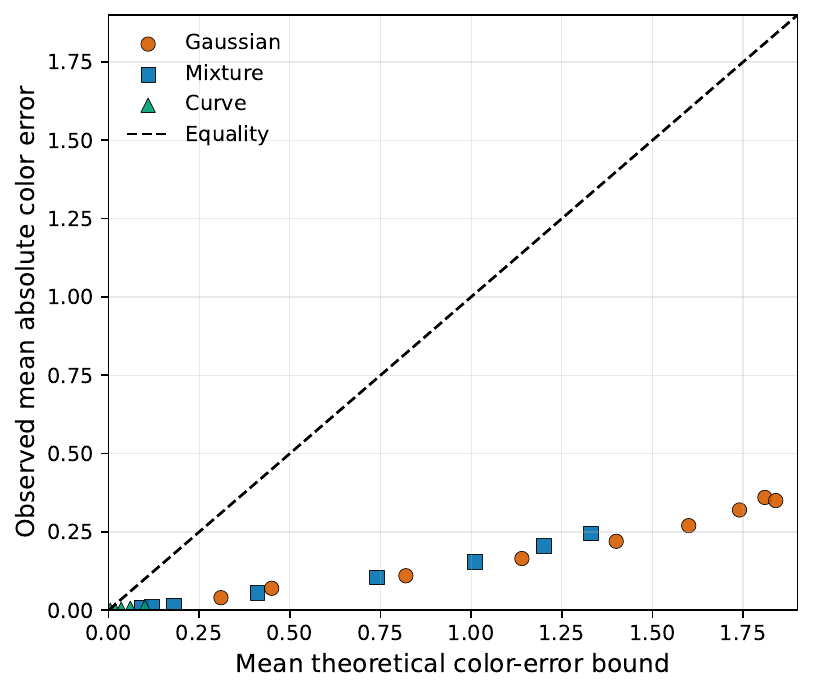}
\caption{Observed mean absolute color error versus the mean theoretical bound for fixed landmark approximate runs.}
\label{fig:color-bound}
\end{figure}

No violations of the mean color error bound were observed in any experiment with fixed landmarks. Figure~\ref{fig:color-bound} compares observed mean absolute color error with the corresponding mean theoretical bound for all approximate fixed landmark runs. All points lie below the equality line. Among configurations with a positive bound, the observed mean absolute error was typically substantially smaller than the bound, with a median error to bound ratio of approximately $0.17$.

\subsection{Landmark Selection and Runtime}

Approximate queries also changed landmark selection. Table~\ref{tab:end-to-end-final} shows the results for the complete Ball Mapper construction at the same representative search effort. In every case the normalized covering radius remained essentially one, as expected for the greedy covering procedure. False negative memberships instead manifested primarily as an increase in the number of selected landmarks and reduced separation. The effect was strongest for Gaussian data and disappeared rapidly for the structured datasets as search effort increased.

\begin{table}[ht]

\centering
\small
\caption{Representative results for the complete Ball Mapper construction across three seeds, reported as mean $\pm$ standard deviation.}
\label{tab:end-to-end-final}
\begin{tabular}{llccc}
\toprule
Dataset & Setting & Landmark ratio & Covering radius/$\varepsilon$ & Separation/$\varepsilon$ \\
\midrule
Gaussian & IVF, \texttt{nprobe}=64 & $1.039\pm0.006$ & $1.0000\pm0.0000$ & $0.839\pm0.023$ \\
Gaussian & HNSW, \texttt{efSearch}=64 & $1.065\pm0.003$ & $1.0000\pm0.0000$ & $0.868\pm0.016$ \\
\addlinespace
Mixture & IVF, \texttt{nprobe}=64 & $1.000\pm0.000$ & $1.0000\pm0.0000$ & $1.000\pm0.000$ \\
Mixture & HNSW, \texttt{efSearch}=64 & $1.021\pm0.002$ & $1.0000\pm0.0000$ & $0.839\pm0.023$ \\
\addlinespace
Curve & IVF, \texttt{nprobe}=64 & $1.000\pm0.000$ & $0.9998\pm0.0002$ & $1.000\pm0.000$ \\
Curve & HNSW, \texttt{efSearch}=64 & $1.000\pm0.001$ & $0.9998\pm0.0002$ & $0.898\pm0.176$ \\
\bottomrule
\end{tabular}
\end{table}

We finally compare these structural effects with runtime.

At $n=20000$, the approximate indexes did not provide a runtime advantage over exhaustive FAISS Flat search. In the representative fixed landmark settings in Table~\ref{tab:representative-fidelity}, the approximate queries required approximately $2.4$--$5.1$ times the Flat query time for mixture and curve data and approximately $4.2$--$5.0$ times the Flat query time for Gaussian data. Configurations that achieved at least $0.99$ edge recall in every seed were slower still: for Gaussian data, IVF \texttt{nprobe}=256 and HNSW \texttt{efSearch}=256 required approximately $8.3$ and $10.0$ times the Flat query time, respectively.

This is an important negative result. The present $n=20000$ experiments validate the fidelity and error propagation behavior of approximate range queries, but they do not establish a computational speed benefit over the highly optimized exhaustive Flat backend. Approximate indexing should therefore be motivated at substantially larger database sizes, under memory constraints, or in settings where exhaustive scanning no longer fits the required query time. The exact FAISS Flat backend remains the preferred implementation at the scales tested here.

\subsection{Interpretation Across Dataset Geometries}

The three geometries consistently ordered themselves by sensitivity to approximation. The isotropic Gaussian dataset was the most difficult: ball overlaps were numerous, and moderate range recall loss caused substantial witness and edge loss. The Gaussian mixture was more stable, while the noisy curve was nearly exact under modest search effort. This dependence on geometry was reproducible across all three seeds.

The boundary margin diagnostics should be interpreted more cautiously. The deterministic boundary results describe perturbations controlled by distance from the radius threshold. IVF and HNSW errors arise instead from candidate omission, and in these experiments ball size covaries with both boundary density and the opportunity for missed candidates. We therefore do not treat the observed boundary margin scatter as an independent empirical validation of the deterministic margin theorem. A controlled perturbation experiment would be needed to isolate that effect.

\section{Limitations and Scope}

The theoretical guarantees are conditional on properties of the range oracle. The additive and multiplicative models do not assert that IVF or HNSW satisfies a universal deterministic error parameter $\delta$ or $\eta$; when such a bound is unavailable, fidelity must be measured against an exact reference. Likewise, the conservative behavior observed for the verified IVF and HNSW configurations is an empirical property of the evaluated implementation and should not be generalized to compressed distance indexes, for which false positives can occur.

The approximation experiments use synthetic Euclidean data at $n=20{,}000$ and CPU execution. They were designed to isolate geometry, witness redundancy, and range query error rather than to establish application specific performance. Real data may exhibit different density, anisotropy, intrinsic dimension, and coloring behavior. The optional hnswlib and cuVS adapters in the software were not benchmarked in this study, and no GPU speed claim is made. Finally, exhaustive FAISS Flat stores vectors in single-precision floating-point format; its search is exhaustive relative to that representation, whereas the exact validation uses double-precision floating-point brute force, BallTree, and cKDTree to make the closed ball mathematical reference explicit.

\section{Conclusion and Future Work}

This work separates the mathematical Ball Mapper construction from the range query backend used to realize its metric balls. Theorem~\ref{thm:exact-oracle-invariance} explains why an exact acceleration can change runtime without changing landmarks, memberships, graph edges, or membership based colorings. Comparisons among the independent brute-force, BallTree, and cKDTree implementations support the consistency of the closed ball computation under the stated numerical convention, while the approximate oracle analysis quantifies how range uncertainty propagates to covering radius, landmark separation, the nerve, edge witnesses, and mean vertex colors.

The approximation study over three seeds yields three main empirical conclusions. Approximation sensitivity is strongly geometry dependent; isotropic Gaussian data are substantially harder than clustered data or data with low intrinsic dimension. Edge stability is governed by witness redundancy, with overlaps supported by multiple witnesses surviving much more reliably than overlaps supported by a single witness. The derived mean color perturbation bound was never violated in the experiments with fixed landmarks.

Approximate indexing did not outperform exhaustive FAISS Flat at $n=20{,}000$. At this scale its value is therefore diagnostic rather than computational: it provides a controlled setting in which to understand fidelity before approximation is used at larger scale. Future work should test substantially larger databases, GPU execution, memory constrained settings, and compressed representations, where a genuine speed--accuracy trade-off is more plausible. Controlled perturbation experiments can separately test boundary margin theory, while real application datasets can determine how local density, witness multiplicity, and auxiliary colorings interact in practice.

\appendix

\section{Backend Implementation Details}\label{app:backend-details}

\subsection{BallTree Implementation}

A \emph{ball tree} \parencite{Omohundro1989} is a rooted binary tree data structure where each node in the tree represents a ball that contains a subset of the data points. A Python implementation of the ball tree data structure is available through scikit-learn \parencite{scikit-learn}.

The scikit-learn implementation uses a balanced binary tree layout. Each node corresponds to a contiguous subset of the indexed observations and stores BallTree bounding information consisting of a centroid and a radius. During recursive construction, scikit-learn chooses the coordinate with the largest spread in the current node and partitions the node indices around the midpoint so that the two children contain approximately equal numbers of observations. Recursion continues until the prescribed leaf size condition is reached; for the default BallTree interface, the leaf size is 40. Changing the leaf size affects construction cost, memory use, and query speed but not the exact result of a range query \parencite{sklearnBallTreeDocs}. This implementation detail is distinct from alternative BallTree construction algorithms, including farthest point pivot schemes described in the classical literature \parencite{Omohundro1989}.

The range query proceeds as follows. Given a query point $q$ and a radius $r$, we start at the root node with center $c$ and radius $R$. 
If the condition
\[
d(q,c)-R>r
\]
is satisfied, then the query ball $B(q,r)$ cannot contain any point stored in that node. Indeed, every stored point $x$ satisfies $d(c,x)\leq R$, so the reverse triangle inequality gives $d(q,x)\geq d(q,c)-R>r$. This does not mean that $B(q,r)$ itself is empty; it means that its intersection with the subset represented by the node is empty. The condition therefore permits the entire subtree to be pruned without examining its individual points.
We then recursively visit each child node and perform the same check. If we reach the leaves, we check the distance of each point in the leaf to determine $B(q,r)$.

Scikit-learn's BallTree supports a range of built-in metrics through its \texttt{valid\_metrics} interface, including Euclidean- and Manhattan type metrics. This flexibility makes ball trees suitable for many metric based implementations of Ball Mapper.

\subsection{$k$-Dimensional Tree Implementation}

A \emph{$k$-dimensional tree}, commonly called a \emph{$k$-d tree}, is a binary tree that organizes observations through recursive coordinate splits \cite{Bentley1975}. Each internal node partitions its observations according to their position relative to a hyperplane perpendicular to a coordinate axis. Repeated partitioning produces smaller subsets, which are stored in the leaves.

For observations in $\mathbb{R}^{D}$, each node is associated with a bounding box
\[
Q=\prod_{j=1}^{D}[a_j,b_j],
\]
where $a_j$ and $b_j$ are lower and upper bounds on the $j$-th coordinates of the observations assigned to that node. For a query point $q=(q_1,\ldots,q_D)$, the minimum Euclidean distance from $q$ to this box is
\[
d_{\min}(q,Q) = \left(\sum_{j=1}^{D}\max\{a_j-q_j,\,0,\,q_j-b_j\}^{2}\right)^{1/2}.
\]
Each term measures the squared distance from a coordinate of the query to the corresponding interval.

A range query with radius $r>0$ can exclude the entire node whenever $d_{\min}(q,Q)>r$. Indeed, every observation $x$ assigned to the node satisfies $d(q,x)\geq d_{\min}(q,Q)$, so none can belong to the closed query ball. Otherwise, the search examines the child nodes. At leaves that have not been excluded, individual distances determine which observations satisfy $d(q,x)\leq r$. Nodes whose bounding boxes lie entirely within the query ball may be accepted without evaluating individual distances.

The implementation uses SciPy’s cKDTree as an independent Euclidean reference for validation \cite{Virtanen2020SciPy}. Its query\_ball\_point method supports Euclidean range queries with $p=2$, while $eps=0$ disables approximate search. This software parameter is distinct from the Ball Mapper radius $\varepsilon$.

\subsection{FAISS Flat Implementation}

For many commonly used metrics, distance computations admit algebraic decompositions that can be exploited for efficiency. We illustrate this for the Euclidean distance. Consider a fixed query point $q$ and the
squared Euclidean distance 
\[
d(q,x) = \left( \sum_{j}\left( q_{j} - x_{j} \right)^{2} \right)^{\frac{1}{2}}.
\] 
The squared Euclidean distance can be expressed as
\[
d^{2}(q,x) = d^{2}(q,0) + d^{2}(x,0) - 2\sum_{j}q_{j}x_{j}.
\]
The term $d^{2}(q,0)$ is constant and computed only once per query. The term $d^{2}(x,0)$ can be precomputed and stored. Consequently, the dominant cost in evaluating lies in the remaining term. This operation requires operations between $q$ and each data point.

FAISS \parencite{douze2025faisslibrary} accelerates this computation by exploiting two mechanisms: single instruction, multiple data (SIMD) processing \parencite{simd2008} and basic linear algebra subprograms (BLAS) \parencite{vandeGeijn2011}. SIMD instructions allow the same arithmetic operation to be applied simultaneously to multiple components of a vector, enabling parallel evaluation of inner products at the hardware level. BLAS is a standardized interface whose implementations are optimized numerical routines for linear algebra. By expressing collections of inner products as matrix--vector or matrix--matrix operations, FAISS leverages highly optimized BLAS routines that are tuned for modern CPU and GPU architectures.

As a result, FAISS reduces the constant factors associated with distance evaluation without altering the underlying algorithmic complexity. In contrast to tree-based data structures, FAISS accelerates the cost per distance computation through low level numerical optimization. Furthermore, FAISS supports parallel execution on modern hardware, which can further speed up computation when sufficient CPU or GPU resources are available.

The implementation studied here uses FAISS for squared Euclidean distance and, optionally, cosine similarity after vector normalization. The theoretical oracle formulation itself is metric agnostic; restrictions on the metric arise from the chosen backend rather than from Ball Mapper.

\section{Extended Exact Benchmark Results}\label{app:extended-benchmark}

\subsection{Memory Usage}

We examine the memory requirements of the algorithms using \emph{peak memory usage}. This refers to the maximum resident memory observed during execution and reported in megabytes (MB). This measure captures the largest memory footprint observed while the algorithm is running, including memory used for data structures, indexing, and intermediate allocations, but excluding memory that is freed before the peak is reached.

\begin{figure}[ht]
\centering
\includegraphics[width=\textwidth]{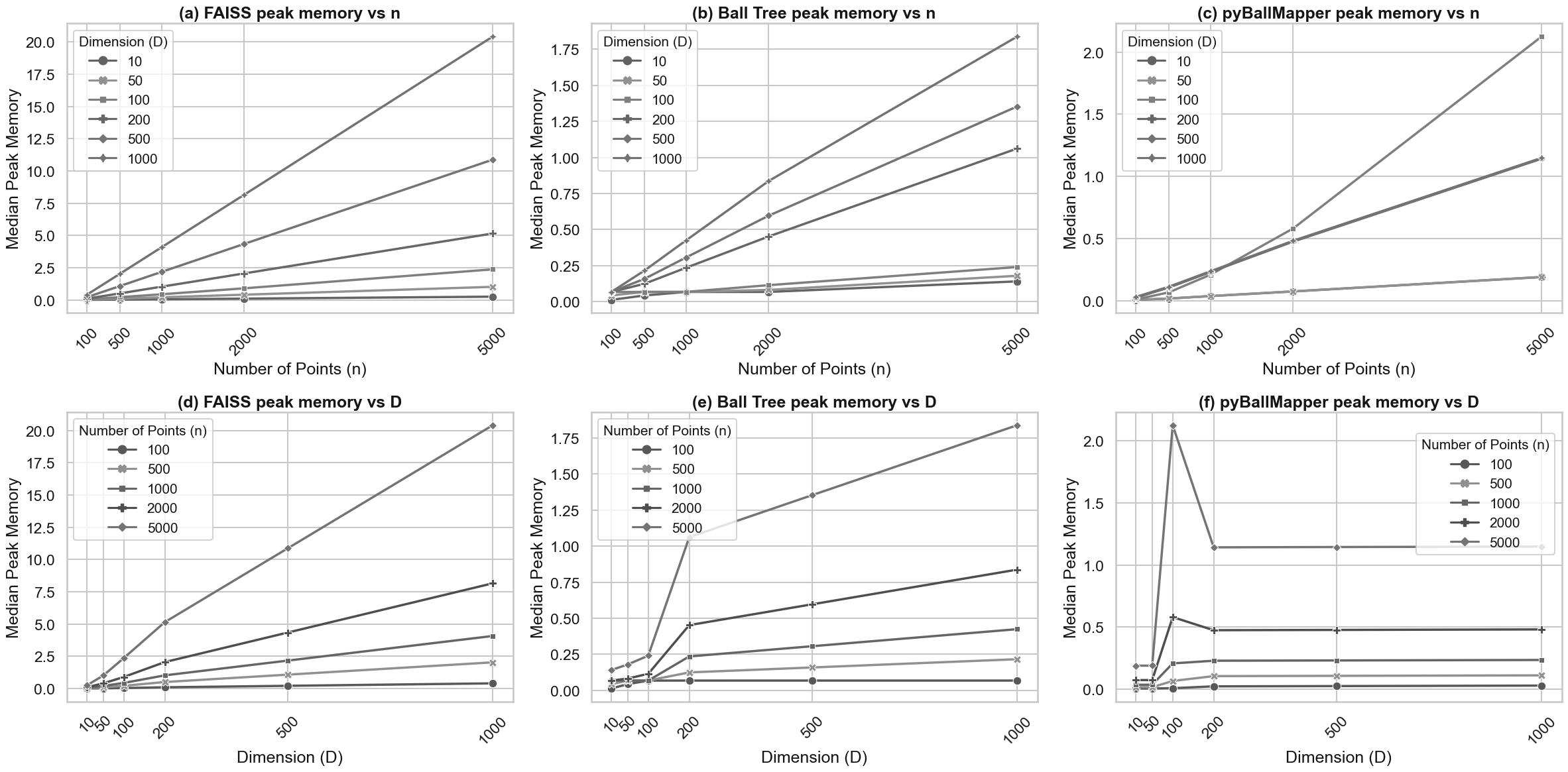}
\caption{Median peak memory usage across varying dataset sizes $n$ and dimensions $D$. Panels (a)--(c) show peak memory as a function of $n$; panels (d)--(f) show peak memory as a function of $D$. Series are distinguished by marker shape and grayscale intensity.}
\label{fig:peakmem}
\end{figure}

Figure~\ref{fig:peakmem} shows the median peak memory usage of the three methods as a function of dataset size $n$. For all algorithms, peak memory increases with $n$, reflecting the cost of storing data-dependent structures that scale with the number of points. FAISS exhibits the highest peak memory usage, particularly for large datasets, consistent with its reliance on dense vector representations and indexing structures optimized for fast search. Ball Tree remains relatively memory-efficient across all dataset sizes, with a more gradual increase in memory usage. pyBallMapper shows lower absolute memory consumption but still exhibits noticeable growth as $n$ increases.

As shown in Figure~\ref{fig:peakmem}, peak memory usage also increases with dimensionality $D$, though the sensitivity varies across methods. FAISS again shows the largest memory footprint at high dimensions, reflecting the cost of storing and processing high-dimensional dense vectors. Ball Tree displays comparatively modest growth in memory usage as $D$ increases. In contrast, pyBallMapper, while maintaining lower overall memory usage, shows increased sensitivity to dimensionality, particularly at intermediate dataset sizes.

Overall, these results highlight a clear trade-off between computational speed and memory usage: FAISS achieves superior runtime performance at the cost of a larger memory footprint, while Ball Tree and pyBallMapper are more memory-efficient but incur higher runtimes in high-dimensional and large-scale settings. The increased memory footprint of FAISS reflects design choices that favor dense data representations and auxiliary indexing structures to maximize computational efficiency.

\subsection{Empirical Scaling Analysis}

The regression below is an \emph{empirical scaling analysis}, not a proof of asymptotic computational complexity. The fitted exponent depends on the finite range of $n$ and $D$, the hardware, implementation, thread count, cache behavior, and the geometry of the sampled data. It should therefore be interpreted together with Proposition~\ref{prop:flat-complexity}, which gives an algorithmic worst-case statement for exhaustive range search.

To summarize finite sample scaling with dataset size, we perform a log--log regression of runtime $T(n)$ as a function of $n$ separately at fixed dimensions $D$. We assume the empirical relationship \[T(n)=kn^p\] over the tested range. Taking logarithms gives 
\[
\log T(n) = \log k + p\log n,
\] 
which expresses $\log T(n)$ as a linear function of $\log n$ at fixed $D$. We estimate this relationship by least squares and call $p$ the \emph{empirical scaling exponent}. Thus, $p=1$ corresponds to linear empirical growth in $n$ over the tested range, while $p=2$ corresponds to quadratic empirical growth. These fitted exponents are descriptive and are not asymptotic complexity guarantees.

The quality of the fit is assessed using the \emph{coefficient of determination} $R^{2}$ defined as
\[
R^{2} = 1 - \frac{\sum_{i}\left( y_{i} - \hat{y_{i}} \right)^{2}}{\sum_{i}\left( y_{i} - \overline{y} \right)^{2}}
\]
where $y_{i}$ are the observed values, $\hat{y_{i}}$ are the predicted values from the regression model, and $\overline{y}$ denotes the mean of $y_{i}$. Higher values of $R^{2}$ indicate a stronger adherence to a power law scaling relationship.

\begin{table}[ht]
\centering
\caption{Empirical scaling exponents and goodness of fit metrics from log--log runtime regression.}
\label{tab:complexity}
\begin{tabular}{llcc}
\toprule
\textbf{Algorithm} & \textbf{$D$} & \textbf{$p$} & \textbf{$R^2$} \\
\midrule
Ball Tree & \multirow{3}{*}{$100$}  & $0.533$ & $0.858$ \\
FAISS & & $0.560$ & $0.852$ \\
pyBallMapper & & $1.157$ & $0.977$ \\
\addlinespace
Ball Tree & \multirow{3}{*}{$200$}  & $1.307$ & $0.961$ \\
FAISS & & $1.159$ & $0.968$ \\
pyBallMapper & & $1.976$ & $1.000$ \\
\addlinespace
Ball Tree & \multirow{3}{*}{$500$}  & $1.609$ & $0.979$ \\
FAISS & & $1.432$ & $0.963$ \\
pyBallMapper & & $2.038$ & $1.000$ \\
\addlinespace
Ball Tree & \multirow{3}{*}{$1000$} & $1.744$ & $0.990$ \\
FAISS & & $1.696$ & $0.968$ \\
pyBallMapper & & $2.045$ & $1.000$ \\
\bottomrule
\end{tabular}
\end{table}

Table~\ref{tab:complexity} reports $p$ and the corresponding $R^{2}$ values for regressions in $n$ at selected fixed dimensions $D$. The fitted exponent increases with $D$ for all three implementations. In the tested range, pyBallMapper approaches an empirical exponent near two at $D\geq200$, whereas FAISS and BallTree remain lower. These values summarize the observed benchmark regime only; they do not replace the worst case statements in Proposition~\ref{prop:flat-complexity}.

\subsection*{Authors' Contributions}

Jay-Anne Bulauan contributed to manuscript drafting, benchmark experiments, and data visualization. John Rick Manzanares contributed to manuscript drafting, algorithm design, mathematical analysis, and software development. Both authors reviewed the manuscript and are responsible for the final content.

\subsection*{Code and Data Availability}

Source code for the range query backends, tests of agreement with the reference implementation, evaluations of approximation errors, experiment scripts, and plotting utilities is publicly available at \url{https://github.com/jhnrckmnznrs/fast-ballmapper}. The experiments in this article use procedurally generated synthetic datasets; the random seeds and generation procedures are specified in the experiment scripts, so no external research dataset is required to reproduce the reported matrix.

\subsection*{Use of Generative AI Tools}

The authors used ChatGPT (OpenAI) to assist with language editing and code review. The authors independently reviewed the resulting text, and take full responsibility for the final manuscript.

\subsection*{Statements and Declarations}

\textbf{Funding.}
This project has received funding from the European Union Horizon Europe research and innovation programme under the Marie Skłodowska-Curie (MSCA) grant agreement No 101120290 (GAP).

\noindent
\textbf{Competing Interests.} The author declares no competing interests.

\bibliographystyle{plainnat}
\bibliography{references}

\begin{thebibliography}{18}
\providecommand{\natexlab}[1]{#1}
\providecommand{\url}[1]{\texttt{#1}}
\expandafter\ifx\csname urlstyle\endcsname\relax
  \providecommand{\doi}[1]{doi: #1}\else
  \providecommand{\doi}{doi: \begingroup \urlstyle{rm}\Url}\fi

\bibitem[Bellman(1957)]{Bellman1957}
Richard~E. Bellman.
\newblock \emph{Dynamic Programming}.
\newblock Princeton University Press, Princeton, NJ, 1957.

\bibitem[Bentley(1975)]{Bentley1975}
Jon~Louis Bentley.
\newblock Multidimensional binary search trees used for associative searching.
\newblock \emph{Communications of the ACM}, 18\penalty0 (9):\penalty0 509--517, 1975.

\bibitem[Borsuk(1948)]{Borsuk1948}
Karol Borsuk.
\newblock On the imbedding of systems of compacta in simplicial complexes.
\newblock \emph{Fundamenta Mathematicae}, 35\penalty0 (1):\penalty0 217--234, 1948.

\bibitem[D{\l}otko(2019)]{dlotko2019ballmapper}
Pawe{\l} D{\l}otko.
\newblock Ball mapper: A shape summary for topological data analysis.
\newblock \emph{arXiv preprint arXiv:1901.07410}, 2019.
\newblock \doi{10.48550/arXiv.1901.07410}.
\newblock URL \url{https://arxiv.org/abs/1901.07410}.

\bibitem[Douze et~al.(2025)Douze, Guzhva, Deng, Johnson, Szilvasy, Mazar{\'e}, Lomeli, Hosseini, and J{\'e}gou]{douze2025faisslibrary}
Matthijs Douze, Alexandr Guzhva, Chengqi Deng, Jeff Johnson, Gergely Szilvasy, Pierre-Emmanuel Mazar{\'e}, Maria Lomeli, Lucas Hosseini, and Herv{\'e} J{\'e}gou.
\newblock The faiss library, 2025.
\newblock URL \url{https://arxiv.org/abs/2401.08281}.
\newblock arXiv version 4.

\bibitem[Flynn(1972)]{simd2008}
Michael~J. Flynn.
\newblock Some computer organizations and their effectiveness.
\newblock \emph{IEEE Transactions on Computers}, C-21\penalty0 (9):\penalty0 948--960, 1972.
\newblock \doi{10.1109/TC.1972.5009071}.
\newblock Foundational reference for SIMD/MIMD computer organization terminology.

\bibitem[Gonzalez(1985)]{GONZALEZ1985293}
Teofilo~F. Gonzalez.
\newblock Clustering to minimize the maximum intercluster distance.
\newblock \emph{Theoretical Computer Science}, 38:\penalty0 293--306, 1985.
\newblock \doi{10.1016/0304-3975(85)90224-5}.

\bibitem[Goto and Van~de Geijn(2008)]{vandeGeijn2011}
Kazushige Goto and Robert~A. Van~de Geijn.
\newblock Anatomy of high-performance matrix multiplication.
\newblock \emph{ACM Transactions on Mathematical Software}, 34\penalty0 (3):\penalty0 12:1--12:25, 2008.
\newblock \doi{10.1145/1356052.1356053}.

\bibitem[Gurnari()]{pyballmapper}
Davide Gurnari.
\newblock pyballmapper.
\newblock URL \url{https://github.com/dioscuri-tda/pyBallMapper}.
\newblock Python implementation of the Ball Mapper algorithm.

\bibitem[J{\'e}gou et~al.(2011)J{\'e}gou, Douze, and Schmid]{Jegou2011PQ}
Herv{\'e} J{\'e}gou, Matthijs Douze, and Cordelia Schmid.
\newblock Product quantization for nearest neighbor search.
\newblock \emph{IEEE Transactions on Pattern Analysis and Machine Intelligence}, 33\penalty0 (1):\penalty0 117--128, 2011.
\newblock \doi{10.1109/TPAMI.2010.57}.

\bibitem[Malkov and Yashunin(2020)]{Malkov2020HNSW}
Yury~A. Malkov and Dmitry~A. Yashunin.
\newblock Efficient and robust approximate nearest neighbor search using hierarchical navigable small world graphs.
\newblock \emph{IEEE Transactions on Pattern Analysis and Machine Intelligence}, 42\penalty0 (4):\penalty0 824--836, 2020.
\newblock \doi{10.1109/TPAMI.2018.2889473}.

\bibitem[{Meta AI Research}(2025)]{faissIndexesWiki}
{Meta AI Research}.
\newblock {Faiss} indexes.
\newblock \url{https://github.com/facebookresearch/faiss/wiki/Faiss-indexes}, 2025.
\newblock GitHub Wiki, accessed 2026-06-19.

\bibitem[{Meta AI Research}(2026)]{faissIndexFlatDocs}
{Meta AI Research}.
\newblock \emph{{Struct faiss::IndexFlat}}.
\newblock Faiss, 2026.
\newblock URL \url{https://faiss.ai/cpp_api/struct/structfaiss_1_1IndexFlat.html}.

\bibitem[Omohundro(1989)]{Omohundro1989}
Stephen~M. Omohundro.
\newblock Five balltree construction algorithms.
\newblock Technical Report TR-89-063, International Computer Science Institute, Berkeley, CA, December 1989.
\newblock URL \url{https://steveomohundro.com/wp-content/uploads/2009/03/omohundro89_five_balltree_construction_algorithms.pdf}.

\bibitem[Pedregosa et~al.(2011)Pedregosa, Varoquaux, Gramfort, Michel, Thirion, Grisel, Blondel, Prettenhofer, Weiss, Dubourg, Vanderplas, Passos, Cournapeau, Brucher, Perrot, and Duchesnay]{scikit-learn}
Fabian Pedregosa, Ga{\"e}l Varoquaux, Alexandre Gramfort, Vincent Michel, Bertrand Thirion, Olivier Grisel, Mathieu Blondel, Peter Prettenhofer, Ron Weiss, Vincent Dubourg, Jake Vanderplas, Alexandre Passos, David Cournapeau, Matthieu Brucher, Matthieu Perrot, and {\'E}douard Duchesnay.
\newblock Scikit-learn: Machine learning in python.
\newblock \emph{Journal of Machine Learning Research}, 12:\penalty0 2825--2830, 2011.
\newblock URL \url{https://jmlr.org/papers/v12/pedregosa11a.html}.

\bibitem[{scikit-learn developers}(2026)]{sklearnBallTreeDocs}
{scikit-learn developers}.
\newblock \emph{{sklearn.neighbors.BallTree}}.
\newblock scikit-learn, 2026.
\newblock URL \url{https://scikit-learn.org/stable/modules/generated/sklearn.neighbors.BallTree.html}.

\bibitem[Sivic and Zisserman(2003)]{SivicZisserman2003}
Josef Sivic and Andrew Zisserman.
\newblock Video google: A text retrieval approach to object matching in videos.
\newblock In \emph{Proceedings of the IEEE International Conference on Computer Vision}, pages 1470--1477, 2003.
\newblock \doi{10.1109/ICCV.2003.1238663}.

\bibitem[Virtanen et~al.(2020)Virtanen, Gommers, Oliphant, Haberland, Reddy, Cournapeau, Burovski, Peterson, Weckesser, Bright, van~der Walt, Brett, Wilson, Millman, Mayorov, Nelson, Jones, Kern, Larson, Carey, Polat, Feng, Moore, VanderPlas, Laxalde, Perktold, Cimrman, Henriksen, Quintero, Harris, Archibald, Ribeiro, Pedregosa, and van Mulbregt]{Virtanen2020SciPy}
Pauli Virtanen, Ralf Gommers, Travis~E. Oliphant, Matt Haberland, Tyler Reddy, David Cournapeau, Evgeni Burovski, Pearu Peterson, Warren Weckesser, Jonathan Bright, St{\'e}fan~J. van~der Walt, Matthew Brett, Joshua Wilson, K.~Jarrod Millman, Nikolay Mayorov, Andrew R.~J. Nelson, Eric Jones, Robert Kern, Eric Larson, C.~J. Carey, {\.{I}}lhan Polat, Yu~Feng, Eric~W. Moore, Jake VanderPlas, Denis Laxalde, Josef Perktold, Robert Cimrman, Ian Henriksen, E.~A. Quintero, Charles~R. Harris, Anne~M. Archibald, Ant{\^o}nio~H. Ribeiro, Fabian Pedregosa, and Paul van Mulbregt.
\newblock Scipy 1.0: Fundamental algorithms for scientific computing in python.
\newblock \emph{Nature Methods}, 17:\penalty0 261--272, 2020.
\newblock \doi{10.1038/s41592-019-0686-2}.

\end{thebibliography}

\end{document}